\documentclass[a4paper,11pt]{article}
\pdfoutput=1 % to allow compilation in JCAP
\usepackage{jcappub}
\usepackage{amsmath}
\usepackage{bm}
\usepackage{graphicx}
\usepackage{enumitem}
\usepackage{geometry}
\usepackage{xspace}
\usepackage{pdflscape}
\usepackage{dsfont}

\usepackage[compat=1.1.0]{tikz-feynman}

\usepackage{mathtools}

\usepackage{relsize}

\allowdisplaybreaks % allows long equations to be displayed over several pages

\makeatletter
\gdef\@fpheader{}
\g@addto@macro\bfseries{\boldmath}
\makeatother

\newcommand{\ie}{{i.e.~}}

\newcommand{\eg}{e.g.~}

\let\oldsqrt\sqrt
\def\sqrt{\mathpalette\DHLhksqrt}
\def\DHLhksqrt#1#2{%
\setbox0=\hbox{$#1\oldsqrt{#2\,}$}\dimen0=\ht0
\advance\dimen0-0.2\ht0
\setbox2=\hbox{\vrule height\ht0 depth -\dimen0}%
{\box0\lower0.4pt\box2}}

\newcommand{\dd}{\mathrm{d}}
\newcommand{\ee}{e}

\newcommand{\lE}{\ell_{\mathrm{E}} }

\newcommand{\sss}[1]{{\scriptscriptstyle{#1}}}

\newcommand{\uPl}{\mathrm{Pl}}

\newcommand{\uc}{\mathrm{c}}

\newcommand{\usssPl}{\sss{\uPl}}

\newcommand{\calP}{\mathcal{P}}

\newcommand{\setI}{\mathbb{I}}

\newcommand{\Ima}{\Im \mathrm{m}\,}

\newcommand{\Mp}{M_\usssPl}

\newcommand{\efolds}{$e$-folds}

\newcommand{\beq}{\begin{equation}}
\newcommand{\eeq}{\end{equation}}
\newcommand{\bea}{\begin{equation}\begin{aligned}}
\newcommand{\eea}{\end{aligned}\end{equation}}

\newlength{\wsingfig}
\newlength{\wdblefig}
\newlength{\wquadfig}
\newlength{\wtriplefig}
\newcommand{\Eq}[1]{Eq.~(\ref{#1})}
\newcommand{\Eqs}[1]{Eqs.~(\ref{#1})}
\newcommand{\Fig}[1]{Fig.~{\ref{#1}}}
\newcommand{\Figs}[1]{Figs.~{\ref{#1}}}
\newcommand{\Ref}[1]{Ref.~{\cite{#1}}}
\newcommand{\Refs}[1]{Refs.~{\cite{#1}}}
\newcommand{\Sec}[1]{Sec.~\ref{#1}}

\newcommand{\App}[1]{Appendix~\ref{#1}}

\newcommand{\deflen}[2]{%      
    \expandafter\newlength\csname #1\endcsname
    \expandafter\setlength\csname #1\endcsname{#2}%
}

\subheader{}

\title{Non Gaussianities from Quantum Decoherence during Inflation}

\author[a]{J\'er\^ome Martin,}
\author[b]{Vincent Vennin}

\affiliation[a]{Institut d'Astrophysique de Paris, UMR
7095-CNRS, Universit\'e Pierre et Marie Curie, 98bis boulevard Arago,
75014 Paris, France}

\affiliation[b]{Laboratoire Astroparticule et Cosmologie, Universit\'e
  Denis Diderot Paris 7, 10 rue Alice Domon et L\'eonie Duquet, 
75013 Paris, France}

\emailAdd{jmartin@iap.fr}
\emailAdd{vincent.vennin@apc.univ-paris7.fr}

\date{today}

\begin{document}
\sloppy

\abstract{Inflationary cosmological perturbations of
  quantum-mechanical origin generically interact with all degrees of
  freedom present in the early Universe. Therefore, they must be viewed
  as an open quantum system in interaction with an environment. This
  implies that, under some conditions, decoherence can take place. The
  presence of the environment can also induce modifications in the
  power spectrum, thus offering an observational probe of
  cosmic decoherence. Here, we demonstrate that this also leads to non
  Gaussianities that we calculate using the Lindblad equation formalism. We show
  that, while the bispectrum remains zero, the four-point correlation
  functions become non-vanishing. Using the Cosmic Microwave
  Background measurements of the trispectrum by the Planck satellite,
  we derive constraints on the strength of the interaction between the
  perturbations and the environment and show that, in some regimes,
  they are more stringent than those arising from the power spectrum.}

\arxivnumber{1805.05609}
\maketitle
\section{Introduction}
\label{sec:intro}
Decoherence is a universal phenomenon that applies to all quantum
systems in interaction with an
environment~\cite{Zurek:1981xq,Joos:1984uk,Schlosshauer:2003zy}. It is
now a well-understood mechanism that has even been observed in the
laboratory~\cite{PhysRevLett.77.4887}. It is usually believed to be
especially relevant for microphysical systems. More recently however,
it has also been shown to be of interest in
cosmology~\cite{Kiefer:1998qe,Bellini:2001jm,Lombardo:2005iz,Kiefer:2006je,
  Martineau:2006ki,Burgess:2006jn,Prokopec:2006fc,
  Sharman:2007gi,Weenink:2011dd, Burgess:2014eoa, Nelson:2016kjm, Rostami:2017akw, Boyanovsky:2018soy}. The
reason is that, according to the theory of cosmic
inflation~\cite{Guth:1980zm,Albrecht:1982wi,Linde:1983gd}, all
structures originate from quantum fluctuations in the
early universe~\cite{Mukhanov:1981xt,Guth:1982ec}. Generically, those fluctuations do not exhaust all degrees of freedom
present at early times and, therefore, if one wants to correctly
describe the evolution of quantum cosmological perturbations, it is
mandatory to consider their interaction with those extra degrees of
freedom. In other words, treating the fluctuations as a closed quantum
system is arguably not the most realistic approach and describing them as
an open quantum system seems to be necessary.

On generic grounds, the evolution of an open quantum system is
described by the Lindblad equation~\cite{Lindblad:1975ef}. The
application of this equation to cosmology has been considered in various
works~\cite{Kiefer:2006je,Burgess:2006jn,Prokopec:2006fc,Martin:2018zbe},
usually in order to study the quantum-to-classical transition that arises from the suppression of 
the off-diagonal terms of the system
density matrix, when expressed in the eigenbasis selected by the form of the interaction with the environment.  More recently, in \Ref{Martin:2018zbe},
it has be shown that the Lindblad equation also implies changes in
the evolution of the diagonal terms themselves, which
give rise to a modification of the power spectrum of the
fluctuations. This entails that decoherence of the cosmological perturbations can
be tested experimentally. Typically, this leads to constraints on the
strength of the interaction between the system and the
environment, which have been worked out in
\Ref{Martin:2018zbe}.

However, decoherence and the Lindblad equation do not only lead to a
modification of the two-point correlation function, but, a priori,
also manifests itself in the higher-order correlators. This is why, in practice,
decoherence or the presence of an interaction between an environment
and the quantum cosmological fluctuations should give rise to non
Gaussianities. The goal of this article is to calculate them and to use
them in order to further constraint cosmic decoherence, thus complementing and possibly improving the results of \Ref{Martin:2018zbe}.

This paper is organised as follows. In \Sec{sec:lindblad}, we
discuss the Lindblad equation and recall how it can be used in the
context of cosmology. In particular, we explain how it can be applied
in order to calculate the modifications of the power spectrum caused
by decoherence. Then, in \Sec{sec:highorder}, we show how the
higher-order correlation functions can be calculated with the help of
the Lindblad equation. In particular, in \Sec{subsec:tri}, we
show that the trispectrum carries information about the
interaction of the perturbation with the environment and in
\Sec{subsec:paramgnl} we calculate the corresponding parameter
$g_{{}_\mathrm{NL}}^{\mathrm{local}}$. In \Sec{sec:constraints},
we use the above result to set new limits on the strength of the
interaction between the system and the environment. Finally, in
\Sec{sec:conclusion}, we present our conclusions. The paper ends
with a series of technical appendices. In \App{sec:applinear}, we
present the calculation of the equations of motion of the correlators
(up to the four point correlation function) in the case of a linear
interaction between the system and the environment. In
\App{sec:appquadratic}, we give the same equations but, this
time, for a quadratic interaction between cosmological perturbations
and an environment. The equations for the three-point correlation
functions are shown in \App{subsec:threequadratic} and for the
four-point correlation functions in \App{subsec:fourquadratic}. In
\App{subsec:mastertri}, we derive a master equation for the
trispectrum and in \App{subsec:solmastertri} we explain how to
solve it.
\section{The Lindblad equation}
\label{sec:lindblad}
Inflationary cosmological perturbations are usually assumed to form an
isolated quantum system. At leading order in perturbation theory, they are then described by the Hamiltonian 
$\hat{H}_v=\frac 12\int \dd ^3 \bm{k}\left[
 \hat{p}_{\bm{k}}\hat{p}_{\bm{k}}^{\dagger}+\omega^2\left(\eta,\bm{k}\right)
\hat{v}_{\bm{k}}\hat{v}_{\bm{k}}^{\dagger}\right]$, where $\hat{v}_{\bm k}(\eta)$ is the Fourier transform of the
Mukhanov-Sasaki variable $v(\eta,{\bm x})$, $\hat{p}_{\bm k}(\eta)$ is its conjugated momentum and $\omega(\eta,{\bm k})$ is the time-dependent frequency of the
Mukhanov-Sasaki variable which reads
$\omega^2(\eta,{\bm k})=k^2-z''/z$ (a prime denotes a derivative with
respect to conformal time) with $z\propto a\sqrt{\epsilon_1}$ where
$\epsilon_1$ is the first Hubble flow parameter,
$\epsilon_1=-\dot{H}/H^2$, $H$ being the Hubble parameter,
$H=\dot{a}/a$ (and a dot denotes a derivative with respect to cosmic
time).

 This is obviously just an approximation since many
other degrees of freedom should generically be present in the early Universe
and interact with them. This means that the
Hamiltonian describing the evolution of the total system is in fact
given by
\begin{align}
\hat{H}=\hat{H}_v\otimes \hat{\setI}_{\mathrm{env}}
+\hat{\setI}_v\otimes \hat{H}_{\mathrm{env}}
+g\hat{H}_{\mathrm{int}},
\end{align}
where $\hat{H}_v$ is the Hamiltonian of cosmological fluctuations given above,
$\hat{H}_{\mathrm{env}}$ is the Hamiltonian of all the other degrees of
freedom, that, in the following, we collectively denotes as the
environment and $\hat{H}_{\mathrm{int}}$ is the interaction Hamiltonian
between them. The strength of this interaction is set by the coupling
constant $g$. If the environment is large compared to the system (namely
contains many more degrees of freedom), then cosmological
perturbations should be viewed as an open quantum system rather than
an isolated one. Moreover, if other reasonable physical conditions are
satisfied,\footnote{\label{footnote:LindbladConditions}Essentially, these conditions are that the environment
evolves on a time scale that is much smaller than that of the system, that the
backreaction of the system on the environment is negligible, and that the
influence of the environment on the system, that is here clearly crucial, can be treated perturbatively. In \Ref{Martin:2018zbe}, it is explained how these conditions can be realised in practice in simple microphysical examples.
}
then one can derive a relatively simple equation of motion
for the density matrix of the system (here the cosmological
perturbations). 
In this case, the master equation controlling $\hat{\rho}_v$, obtained
from the full density matrix by tracing out the environmental degrees
of freedom, is the Lindblad equation~\cite{Lindblad:1975ef}
\begin{align}
\label{eq:Lindblad}
\frac{{\rm d}\hat{\rho}_v}{{\rm d}\eta}=-i\left[\hat{H}_v,\hat{\rho}_v\right]
-\frac{\gamma}{2}\int {\rm d}^3{\bm x}\, {\rm d}^3{\bm y}\, 
C_R({\bm x},{\bm y})\left[\hat{A}({\bm x}), \left[\hat{A}({\bm y}), 
\hat{\rho}_v\right]\right]\, .
\end{align}
In this expression, we have assumed a local interaction of the form
$\hat{H}_{\mathrm{int}}(\eta)=\int {\rm d}^3{\bm x}\, \hat{A}(\eta, {\bm
  x})\otimes \hat{R}(\eta,{\bm x})$, $\eta$ being the conformal time, and defined the environmental
correlation function by
$C_R({\bm x},{\bm y})\equiv \left\langle \hat{R}(\eta,{\bm
    x})\hat{R}(\eta,{\bm y})\right\rangle$.
The parameter $\gamma $ is given by $\gamma =2g^2\eta_{\mathrm{c}}$,
where $\eta_{\mathrm{c}}$ is the autocorrelation time of the
environment. In a cosmological context, $\gamma $ is generically a
function of time and can be written as $\gamma=\gamma_*(a/a_*)^p$, where
$p$ is a free index and a star refers to a reference time.

Despite its simplicity the Lindblad equation remains difficult to
solve. The only situation where an exact solution is available is when
the interaction is linear in the Mukhanov-Sasaki variable, $\hat{A}=\hat{v}$. Indeed, it
was shown in \Ref{Martin:2018zbe} that, in that case, the
density matrix of the system remains Gaussian, and an explicit expression was derived. However, knowing the
entire density matrix is not necessarily mandatory if one is only interested in some of the system's correlation functions.
From the Lindblad equation indeed, one can
derive an equation of evolution for the quantum expectation value $ \langle \hat{O}\rangle
=\mathrm{Tr}(\hat{\rho}_v\hat{O})$ of an arbitrary operator $\hat{O}$ acting in the Hilbert space
of the system. This equation, written in Fourier space, reads
\begin{align}
\label{eq:Lindblad:mean:Fourrier}
\frac{\dd \left\langle \hat{O}\right\rangle}
{\dd  \eta}&=
\left\langle\frac{\partial \hat{O}}{\partial \eta}\right\rangle
-i \left\langle \left[\hat{O},\hat{H}_{v}\right] \right\rangle
%\nonumber \\ &
-\frac{\gamma}{2}(2\pi)^{3/2}\int\dd ^3\bm k  
\, \tilde{C}_{R}\left({\bm k}\right)
\left\langle\left[\left[\hat{O},\hat{A}_{\bm k}\right],
\hat{A}_{-{\bm k}}\right]\right\rangle\, .
\end{align}
In the above formula, we have assumed that the environment is placed
in a statistically homogeneous configuration such that
$C_R(\bm{x},\bm{y})=C_R(\bm{x}-\bm{y})$. As a consequence, the Fourier
transform of the environmental correlation function is performed against $\bm{x}-\bm{y}$.

Endowed with \Eq{eq:Lindblad:mean:Fourrier}, one can then
calculate various correlators of the system. For the mean value of the
operators $\hat{v}_{\bm k}$ and $\hat{p}_{\bm k}$, one always has
\begin{align}
\label{eq:linearlinear}
  \frac{\dd  \left\langle \hat{v}_{\bm k}\right\rangle}{\dd  \eta } =
  \left \langle \hat{p}_{\bm k}\right \rangle\, , \qquad 
  \frac{\dd  \left\langle \hat{p}_{\bm k} \right\rangle}{\dd  \eta } =
  -\omega^2(\eta, {\bm k})\left \langle \hat{v}_{\bm k}\right \rangle \, .
\end{align}
Combining these two equations, one obtains $\langle \hat{v}_{\bm k}\rangle''+\omega^2 \langle \hat{v}_{\bm k}\rangle=0$, \ie $\left \langle \hat{v}_{\bm k}\right \rangle $
follows the classical equation of motion (the so-called Mukhanov-Sasaki equation). 

One can then go on and determine the two-point correlation
functions
$\langle \hat{O} \rangle = \langle \hat{O}_{{\bm k}_1} \hat{O}_{{\bm
    k}_2}\rangle$,
with $\hat{O}_{{\bm k}_i}=\hat{v}_{\bm{k}_i}$ or
$\hat{p}_{\bm{k}_i}$. This was done explicitly in
\Ref{Martin:2018zbe}. These correlators, due to statistical
homogeneity, can always be written as
$\left\langle \hat{O}_{{\bm k}_1}\hat{O}'_{{\bm k}_2}\right \rangle
=P_{OO'}(\bm{k}_1)\, \delta ({\bm k}_1+{\bm k}_2)$.
Clearly, one is mostly interested in $P_{vv}$ since it is directly
related to the power spectrum of curvature perturbations that one can probe observationally. It was shown to obey a third-order differential
equation given by
\begin{align}
\label{eq:thirdv}
  P_{vv}'''\left(\eta,\bm{k}\right)+4\, \omega^2(\eta,{\bm k})P_{vv}'\left(\eta,\bm{k}\right)
+4\, \omega'(\eta,{\bm k})\omega (\eta,{\bm k})P_{vv}\left(\eta,\bm{k}\right)=S_n(\eta,{\bm k})\, ,
\end{align}
where $S_n$ is a source function that depends on the order $n$
of the interaction $\hat{A}=\hat{v}^n(\eta,{\bm x})$. In
\Ref{Martin:2018zbe}, the solution to this equation was found to be
\begin{align}
\label{eq:exactPvv}
P_{vv}(\eta,{\bm k})=v_{\bm k}(\eta)v_{\bm k}^*(\eta)
+\frac{2}{-W^2}\int_{-\infty}^{\eta}
\dd\eta'S_n(\eta',{\bm k})\, \Ima^2
\left[v_{\bm k}\left(\eta'\right)v_{\bm k}^*\left(\eta\right)\right]\, ,
\end{align}
where $v_{\bm k}$ is a solution of the Mukhanov-Sasaki equation mentioned above and $W=v^*_{\bm k} v_{\bm k}'  - v'^*_{\bm k}v_{\bm k}$ is its conserved Wronskian. Different solutions $v_{\bm k}$ to the Mukhanov-Sasaki equation correspond to different initial conditions. For instance, if one wishes to start in the Bunch-Davies vacuum, one needs to take the solution for which, in the sub-Hubble regime, $v_{\bm k}=\ee^{-ik\eta}/\sqrt{2k}$ and $W=i$. In \Eq{eq:exactPvv}, the first term corresponds to the standard result and the second term is a correction caused by the interaction with the environment. 
In \Ref{Martin:2018zbe}, the source function was calculated
using different techniques, either directly (for $n=1$ and $n=2$) or
diagrammatically (for arbitrary $n$), leading to an explicit
calculation of the modification of the power spectrum originating from
quantum decoherence. As mentioned before, our goal in this paper is to extend this analysis to higher-order
correlators.

Before turning to this question, let us mention that the amount of decoherence the system undergoes has also been studied in \Ref{Martin:2018zbe}. It can be characterised by the decoherence parameter $\delta_{\bm{k}}$ that measures the deviation from a pure state through $\mathrm{Tr}\left(\hat{\rho}_{\bm k}{}^2\right) =(1+4\delta_{\bm k})^{-1/2}$, where $\hat{\rho}_{\bm k}$ is the density matrix of the system traced over all wavenumbers but $\bm{k}$. It can be expressed in terms of the same source function that appears in the correction to the power spectrum,
\bea
\label{eq:quad:delta:intPvv:generic}
\delta_{\bm{k}}\left(\eta\right) = \frac{1}{2} \int_{-\infty}^\eta S_n\left(\eta',\bm{k}\right) P_{vv}\left(\eta',\bm{k}\right) \dd \eta'\, .
\eea
The system is said to have decohered when $\delta_{\bm{k}} \gg 1$. In that case, the off-diagonal element of the density matrix that carries the phase information between two realisations $v_{\bm{k}}+\Delta v_{\bm{k}}$ and $v_{\bm{k}}-\Delta v_{\bm{k}}$ is suppressed by a factor $\ee^{-\delta_{\bm{k}}/2}$ if the two realisations are separated by the typical fluctuation amplitude $\Delta v_{\bm{k}}=\sqrt{P_{vv}}(\bm{k})$.

\section{Higher-order correlation functions}
\label{sec:highorder}

Let us first notice that the case of a linear interaction in the
Mukhanov-Sasaki variable is peculiar. Indeed, as already mentioned, it
was shown in \Ref{Martin:2018zbe} that the Lindblad equation
can be solved exactly and that the density matrix of the system
remains Gaussian. Therefore, although the power spectrum receives
corrections (the system remains Gaussian but the variance
of the Gaussian needs not to be the same), the higher-order
correlators remain exactly zero. These considerations are confirmed
explicitly in \App{sec:applinear} where we show that
the three- and four-point connected correlators vanish. This can also be understood diagrammatically by noticing that one cannot build a connected diagram with more than two external system's propagators and vertices that only involve one system's propagator and one environment's propagator.

For this reason, one needs to consider more complicated, non-linear
types of interaction. In the following, we study quadratic
interactions for which the operator $\hat{A}$ is given by
$\hat{v}^2(\eta, {\bm x})$. The general situation
$\hat{A}=\hat{v}^n(\eta, {\bm x})$ is technically more
complicated but can in principle be addressed along similar lines and we comment on this case in the conclusion. Moreover,
we restrict our considerations to the three-point (bispectrum)
and four-point (trispectrum) correlation functions since higher connected correlators vanish for quadratic interactions. The details of the
calculations are presented in \App{sec:appquadratic}.

Let us start with the bispectrum. In
\App{subsec:threequadratic}, we have calculated the
three-point correlators
$\langle \hat{O} \rangle = \langle \hat{O}_{{\bm k}_1} \hat{O}_{{\bm
    k}_2}\hat{O}_{{\bm k}_3}\rangle $
with $\hat{O}_{{\bm k}_i} =\hat{v}_{\bm{k}_i}$ or
$\hat{p}_{\bm{k}_i}$. At leading order in $\gamma $ (an assumption
which is made in the derivation of the Lindblad equation, see footnote~\ref{footnote:LindbladConditions}), the general
solution is such that these correlators remain zero. This means that
the bispectrum vanishes at leading order in $\gamma$ even if the interaction is quadratic. This can again be understood diagrammatically by noticing that one cannot draw a connected diagram with three external system's legs, since vertices involve two system's propagators and one environment's propagator.

\subsection{The trispectrum}
\label{subsec:tri}

As a consequence, non Gaussianities can only be encountered by considering higher-order correlation functions. In
\App{subsec:fourquadratic}, we have studied the four-point
correlators, namely
$\langle \hat{O} \rangle = \langle \hat{O}_{{\bm k}_1} \hat{O}_{{\bm
    k}_2}\hat{O}_{{\bm k}_3}\hat{O}_{{\bm k}_4}\rangle $
with $\hat{O}_{{\bm k}_i} =\hat{v}_{\bm{k}_i}$ or
$\hat{p}_{\bm{k}_i}$. This time, non Gaussianities are present and,
therefore, interaction with an environment is responsible for a
non-vanishing trispectrum that we calculate in this section. It is
interesting to notice that decoherence is one of the rare examples
where the bispectrum is perturbatively suppressed compared to the trispectrum, which therefore contains the relevant signal.

Let us discuss the non linearity-parameters characterising the
trispectrum. They can be defined in various ways, but to make easy contact
with observations, let us introduce the local $g_{{}_\mathrm{NL}}$
parameter as in \Refs{Ade:2015ava, Martin:2015dha}, defined by
\begin{align} 
\label{eq:deftri}
\left\langle
\zeta_{\bm{k}_1}\zeta_{\bm{k}_2}\zeta_{\bm{k}_3}\zeta_{\bm{k}_4}\right\rangle_\uc
=&
\frac{54}{25}g_{{}_\mathrm{NL}}^{\mathrm{local}}
\left[P_\zeta\left(\bm{k}_1\right)P_\zeta\left(\bm{k}_2\right)
P_\zeta\left(\bm{k}_3\right)+3\,\mathrm{permutations}\right]
\nonumber \\ & \times
\delta\left(\bm{k}_1+\bm{k}_2+\bm{k}_3+\bm{k}_4\right)\,
, 
\end{align} 
see Eq.~(64) in \Ref{Ade:2015ava}, where the index ``$\mathrm{c}$''
denotes the connected part of the correlator. Cosmic Microwave
Background (CMB) measurements constraint
$g_{{}_\mathrm{NL}}^\mathrm{local}=(-9.0+7.7)\times 10^4$ at the
one-sigma level~\cite{Ade:2015ava}. Since the curvature perturbations
$\zeta_{\bm k}$ is related to the Mukhanov-Sasaki variable $v_{\bm k}$
by $\zeta_{\bm{k}}=v_{\bm{k}}/(a\sqrt{2\epsilon_1}\Mp)$, the
calculation of the trispectrum boils down to the calculation of the
four-point correlation function of $\hat{v}_{\bm k}$ as done in
\App{subsec:fourquadratic}. The evolution equations for the four-point correlators yield a single differential equation of order sixteen for $\langle \hat{v}_{{\bm k}_1}\hat{v}_{{\bm k}_2}\hat{v}_{{\bm k}_3}
\hat{v}_{{\bm k}_4}\rangle_\uc$,
which is not especially illuminating. However, if one assumes that the
wavenumbers $\bm{k}_1$, $\bm{k}_2$, $\bm{k}_3$ and $\bm{k}_4$ all
have the same modulus $k$, or, in other words, form a losange (the semi-angle 
at the top of which -- it is the angle between $\bm{k}_1$ and
$\bm{k}_1+\bm{k}_2$ -- is denoted $\alpha$ in the following), then
the system of equations can be reduced to a differential equation
of order five only. This equation is derived in
\App{subsec:mastertri} and reads
\begin{align}
  \biggl\lbrace
  \frac{\dd^5}{\dd\eta^5}
  & +20\, \omega^2\frac{\dd^3}{\dd\eta^3}
  +60\, \omega\, \omega'\frac{\dd^2}{\dd\eta^2}
  +\left[64\, \omega^4+18\left(\omega^2\right)''\right]\frac{\dd}{\dd\eta}
\nonumber \\ &
  +\left[128 \, \omega^3\omega'+4\left(\omega^2\right)'''\right]
  \biggr\rbrace
  \left\langle \hat{v}_{{\bm k}_1}\hat{v}_{{\bm k}_2}\hat{v}_{{\bm k}_3}
\hat{v}_{{\bm k}_4}\right\rangle_\uc= \mathfrak{S}(\eta,{\bm k},\alpha),
\label{eq:mastereq}
\end{align}
where, at leading order in $\gamma$, the source function
$\mathfrak{S}(\eta,{\bm k},\alpha)$ is given by
\begin{align}
\label{eq:sourcetri}
\mathfrak{S}(\eta, {\bm k},\alpha) =&\frac{32}{\left(2\pi\right)^{{3/2}}}
\left(
P_{vv}^2(k)\left[\gamma\, \tilde{C}_R\left(2k \cos\alpha\right)
+\gamma\, \tilde{C}_R\left(2k \sin\alpha\right)+\gamma\, \tilde{C}_R(0)\right]''
\right. \nonumber \\ & \left.
+7P_{vv}(k)\left[P_{vp}(k)+P_{pv}(k)\right]
\left[\gamma\, \tilde{C}_R\left(2k \cos\alpha\right)
+\gamma\, \tilde{C}_R\left(2k \sin\alpha\right)+\gamma\, \tilde{C}_R(0)\right]'
\right. \nonumber \\ & \left.
+2\left\{4\left[P_{vp}(k)+P_{pv}(k)\right]^2+5 P_{vv}(k)P_{pp}(k)
-3\omega^2(k)P_{vv}^2(k)\right\}
\right. \nonumber \\ & \left.
\times\left[\gamma\, \tilde{C}_R\left(2k \cos\alpha\right)
+\gamma\, \tilde{C}_R\left(2k \sin\alpha\right)+\gamma\, \tilde{C}_R(0)\right]
\right) \delta\left(\bm{k}_1+\bm{k}_2+\bm{k}_3+\bm{k}_4\right)\, .
\end{align}
The analogy between \Eqs{eq:mastereq} and~(\ref{eq:thirdv}) is
striking. One can even conjecture that any correlator must obey a linear
differential equation with a source term that describes the interaction with the environment. The differential operator
should be independent of the interaction and its order $2^m$ is determined by the order $m$ of the correlator and possibly reduced by the underlying symmetries
(in our case, restricting to equilateral configurations has reduced the order
from sixteen to five). On the other hand, the source depends on the
interaction and is more complicated to calculate for
higher-order interactions.

The analogy between \Eqs{eq:mastereq} and~(\ref{eq:thirdv})
becomes more explicit if one notices that \Eq{eq:mastereq}
admits an exact solution that has the same structure as
\Eq{eq:exactPvv} and reads
\begin{align}
\label{eq:soltri}
\left\langle \hat{v}_{{\bm k}_1}\hat{v}_{{\bm k}_2}\hat{v}_{{\bm k}_3}
\hat{v}_{{\bm k}_4}\right\rangle_\uc = \frac{2}{3 W^4}
\int_{-\infty}^\eta \dd\eta'\mathfrak{S}(\eta',{\bm k},\alpha)\Ima^4
\left[v_{\bm k}\left(\eta'\right) v_{\bm k}^*\left(\eta\right)\right]\, .
\end{align}
As before, $v_{\bm k}$ is a solution of the Mukhanov-Sasaki equation and
$W$ is its (preserved) Wronskian. This means that non
Gaussianities caused by decoherence can be calculated analytically
exactly, which is a priori highly non trivial. Using the definition of
the trispectrum~(\ref{eq:deftri}), the
$g_{{}_\mathrm{NL}}^{\mathrm{local}}$ parameter can be expressed as
\begin{align}
\label{eq:gnl}
g_{{}_\mathrm{NL}}^{\mathrm{local}}=&
\frac{25}{324}\frac{(a\sqrt{2\epsilon_1}\Mp)^{2}}
{P_{vv}^3\left(\bm{k}\right)}
\int_{-\infty}^\eta \dd\eta' \mathfrak{S}\left(\eta',{\bm k},\alpha\right)
\Ima^4\left[v_{\bm k}\left(\eta'\right) v_{\bm k}^*\left(\eta\right)\right]\, .
\end{align}

\subsection{Computing the $g_{{}_\mathrm{NL}}$ parameter}
\label{subsec:paramgnl}

In order to calculate the trispectrum, one needs to perform the
integral in \Eq{eq:gnl}, which, given the
formula~(\ref{eq:sourcetri}), requires the knowledge of the
environmental correlation function and its Fourier transform $\tilde{C}_R({\bm k})$. Following
\Ref{Martin:2018zbe}, we assume that it takes the form 
\begin{align}
\label{eq:Ck:appr}
C_R\left(\bm x , \bm y\right) = \bar{C}_R\, \Theta\left(\frac{a\vert \bm x - \bm y \vert}{\lE}\right)\, ,
\end{align}
where $\Theta(z)=1$ if $z<1$ and zero otherwise. Here
$\lE $ is the correlation length of the environment. If it is equal to the time scale over which the environment varies, since the system typically varies over Hubble times, $\lE$ needs to be much smaller than the Hubble length $(H\lE\ll 1)$, see footnote~\ref{footnote:LindbladConditions}. The Fourier transform of the environmental correlation function can be then approximated by $\tilde{C}_R({\bm k}) = \sqrt{2/{\pi}} \bar{C}_R\lE^3/({3a^3} )
\Theta(k\lE/{a})$.
It is worth
noticing that the above choice does not influence much the final
result, the only important feature being the existence of a preferred
scale $\lE$ beyond which the correlation function
vanishes. In practice, it means that the lower bound in the integral~(\ref{eq:gnl}) becomes finite.

The calculation of the solution~(\ref{eq:soltri}) is performed in
\App{subsec:solmastertri}. There, it is shown that the
amplitude of the trispectrum is controlled by the dimensionless
quantity $\sigma_\gamma\equiv\bar{C}_R\lE^3\gamma_*/a_*^3$ that parametrises the strength of the interaction with the environment. Approximated formulas can be derived, which depend on where the integral~(\ref{eq:gnl}) receives its main contribution from. Let us give these expressions in the limit where $-k\eta\ll 1$ (the trispectrum is evaluated on super-Hubble scales, where observable modes lie at the end of inflation) and $H\lE\ll 1$ for the reason mentioned above.  

If $p<4$, one finds 
\begin{align}
\label{eq:gNL:pLT4}
\left.g_{{}_\mathrm{NL}}^{\mathrm{local}}\right\vert_{p<4}=&
\frac{25\sigma_\gamma}{1296(4-p)\pi^4\calP_\zeta(k_*)} 
\left(\frac{k}{k_*}\right)^{p-3}\left[\left(2\cos\alpha H\lE\right)^{p-4}
+\left(2\sin\alpha H\lE\right)^{p-4}
\nonumber \right. \\ & \left.
-2-2\left(p-4\right)\left(\ln 4 + \gamma_{{}_\mathrm{E}}\right)\right],
\end{align}
where $\gamma_{{}_\mathrm{E}}$ is the Euler number, $k_*$ is the comoving scale that crosses out the Hubble radius at the reference time introduced above, and $\calP_\zeta(k)\equiv k^3 P_{\zeta}/(2\pi^2)$.\footnote{In \Eqs{eq:gNL:pLT4}-(\ref{eq:gNL:pGT6}), since $g_{{}_\mathrm{NL}}^{\mathrm{local}}$ is proportional to $\gamma$ through the parameter $\sigma_\gamma$, at leading order in $\gamma$ it is enough to evaluate the power spectrum in the free theory [\ie including the first term in the right-hand side of \Eq{eq:exactPvv} only] where one has $\calP_\zeta=H^2/(8\pi^2\Mp^2\epsilon_1)\simeq 2.2\times 10^{-9}$ at the pivot scale $k_*=0.05\,\mathrm{Mpc}^{-1}$~\cite{Ade:2015lrj}.} In this case, the main contribution to the integral~(\ref{eq:gnl}) comes from its lower bound and this is why the environmental correlation length $\lE$ explicitly appears. The case $p=4$ is singular and
needs to be treated separately, leading to
\begin{align}
\left.g_{{}_\mathrm{NL}}^{\mathrm{local}}\right\vert_{p=4}=&
\frac{25\sigma_\gamma}{648\pi^4\calP_\zeta(k_*)} 
\frac{k}{k_*} \left[
 \gamma_{{}_\mathrm{E}}+\ln(2)
-\ln\left(\sqrt{\vert\cos\alpha\sin\alpha\vert} H\lE\right)
 \right]\, .
\end{align}

\begin{figure}[t]
\begin{center}
\includegraphics[width=0.48\textwidth]{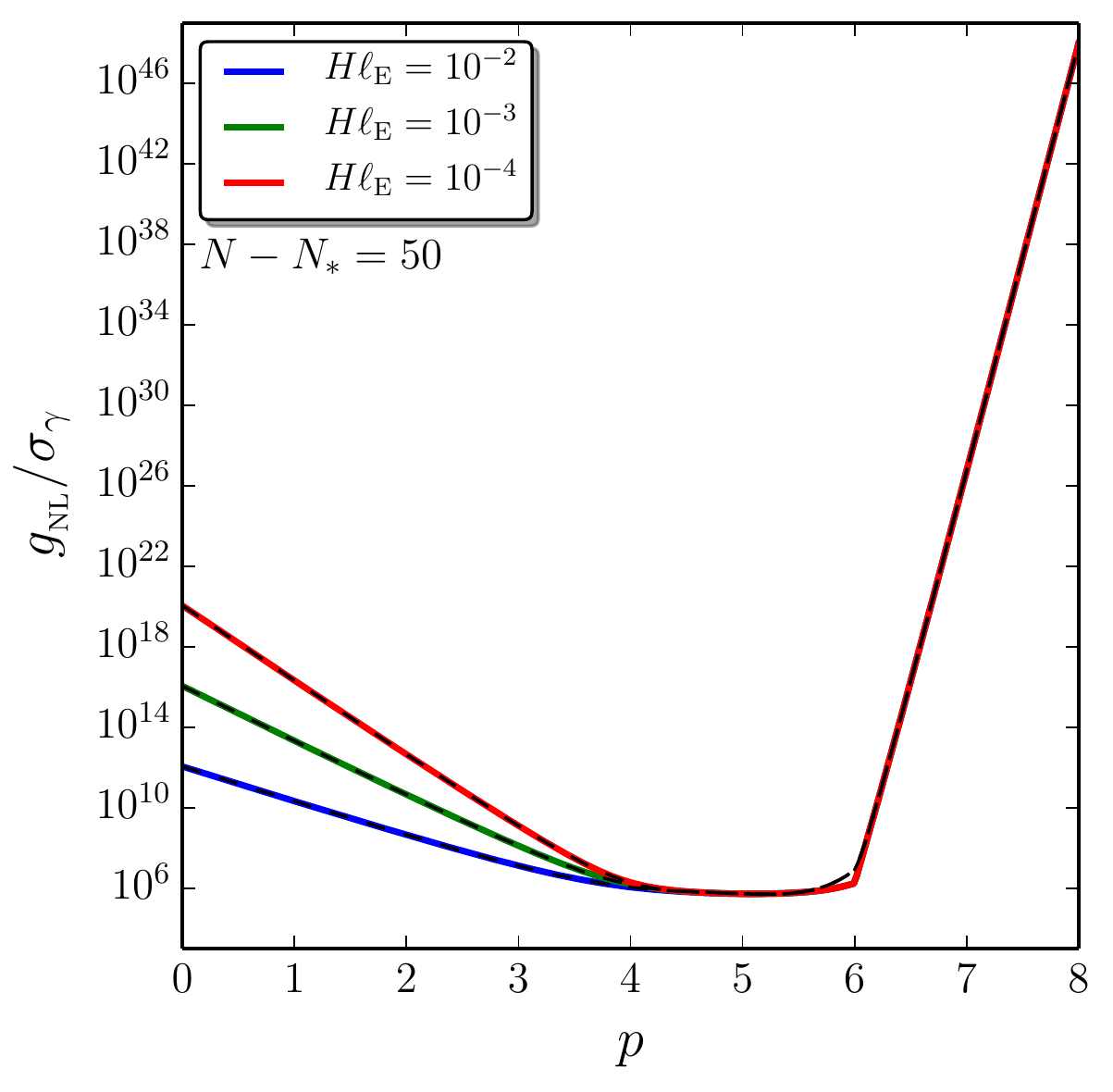}
\includegraphics[width=0.48\textwidth]{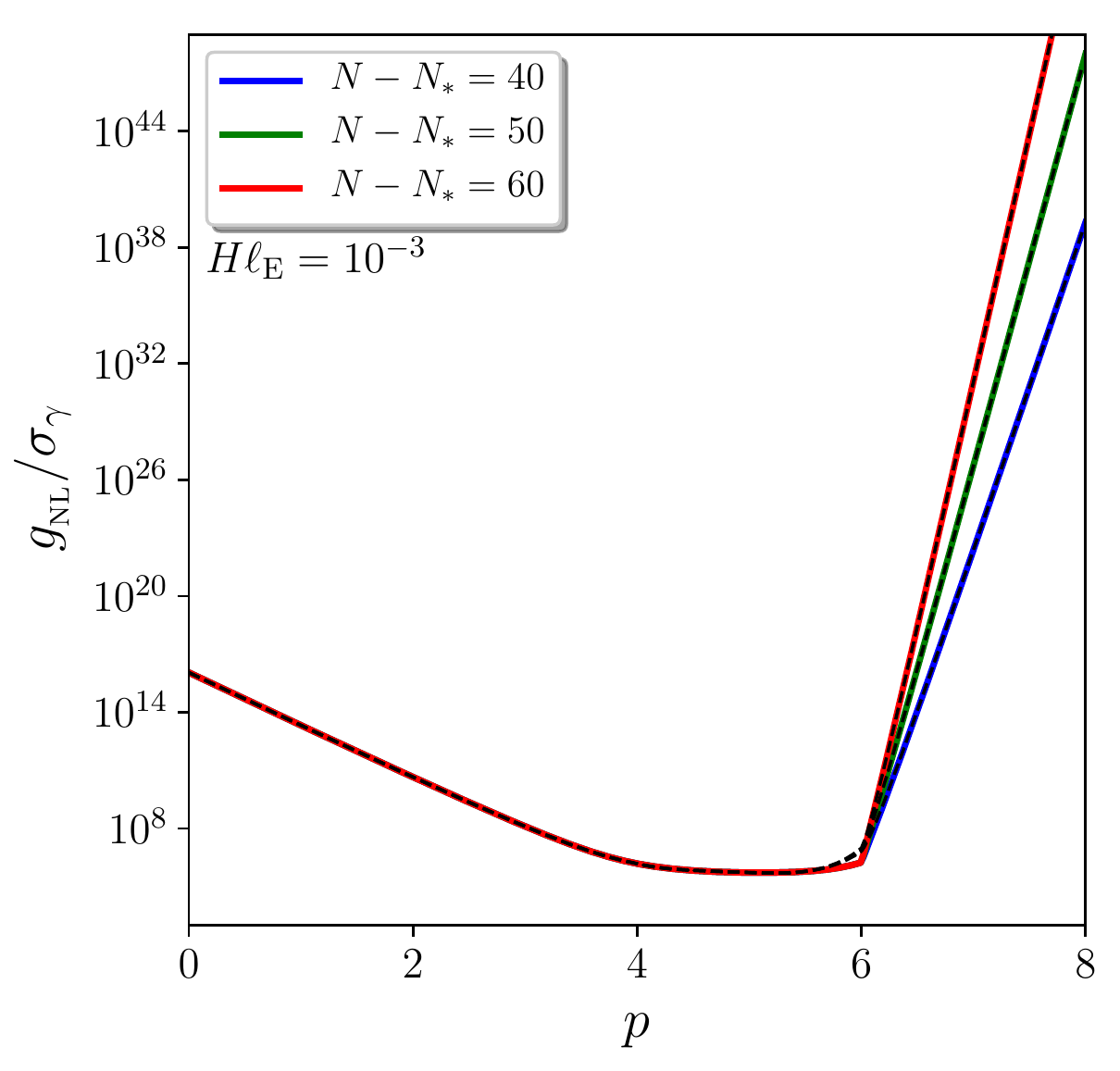}
\caption{Non Gaussianity parameter $g_{{}_\mathrm{NL}}$ rescaled by $\sigma_{\gamma}$ and as a function of $p$, for $N-N_*=50$ and a few values of $H\lE$ (left panel), and for $H\lE=10^{-3}$ and a few values of $N-N_*$ (right panel). The other parameters have been fixed to $k/k_*=1$ and $\alpha=\pi/4$. The coloured lines correspond to a numerical integration of \Eq{eq:gnl} while the black dashed lines correspond to our analytical approximations~(\ref{eq:gNL:pLT4})-(\ref{eq:gNL:pGT6}).}
\label{fig:gNL:p}
\end{center}
\end{figure}
If $4<p<6$, one has
\begin{align}
\left.g_{{}_\mathrm{NL}}^{\mathrm{local}}\right\vert_{4<p<6}=&
\frac{25\sigma_\gamma}{243\pi^4\calP_\zeta(k_*)} \mathcal{O}(1)
\left(\frac{k}{k_*}\right)^{p-3}\, .
\end{align}
In this case, the integral~(\ref{eq:gnl}) receives its dominant contribution from intermediate values $-k\eta'\sim 1$, which makes 
the overall amplitude more difficult to predict and explains why the result is given up to an order of one constant.\footnote{\label{footnote:interpolation}This constant can be calculated exactly in the case of half-integer values of $p$. For $p=4.5$, one finds $103\sqrt{\pi/2}/90$, for $p=5$ one finds $7\pi/20$ and for $p=5.5$ one finds $194\sqrt{2\pi}/385$. In practice in \Fig{fig:gNL:p}, we use a polynomial interpolator between these values.} The case $p=6$ is, as the case $p=4$, peculiar and gives rise to
\begin{align}
\left.g_{{}_\mathrm{NL}}^{\mathrm{local}}\right\vert_{p=6}=&
\frac{25\sigma_\gamma}{243\pi^4\calP_\zeta(k_*)} 
\left(\frac{k}{k_*}\right)^{3} 
\left[1-\frac{\gamma_{{}_\mathrm{E}}+\ln 4}{3}
-\frac{1}{3} \ln\left(\frac{k}{k_*}\right) +\frac{N-N_*}{3} 
\right]\, .
\end{align}

\begin{figure}[t]
\begin{center}
\includegraphics[width=0.48\textwidth]{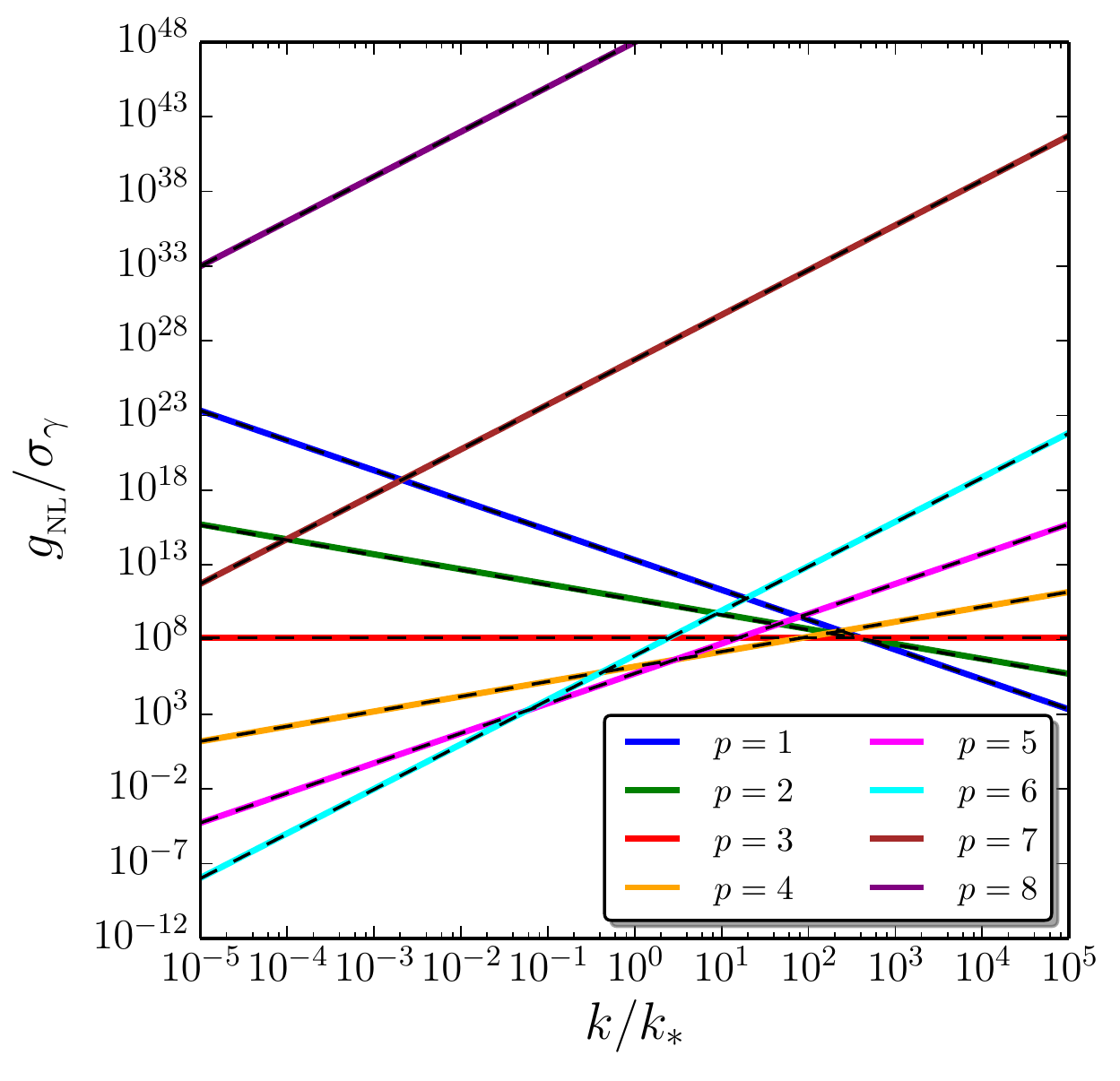}
\caption{Non Gaussianity parameter $g_{{}_\mathrm{NL}}$ rescaled by $\sigma_{\gamma}$ and as a function of $k/k_*$ and a few values of $p$, for $N-N_*=50$, $H\lE=10^{-3}$ and $\alpha=\pi/4$. The coloured lines correspond to a numerical integration of \Eq{eq:gnl} while the black dashed lines correspond to our analytical approximations~(\ref{eq:gNL:pLT4})-(\ref{eq:gNL:pGT6}).}
\label{fig:gNL:k}
\end{center}
\end{figure}
Finally, if $p>6$, one finds
\begin{align}
\label{eq:gNL:pGT6}
\left. g_{{}_\mathrm{NL}}^{\mathrm{local}}\right\vert_{p>6}=&
\frac{50\sigma_\gamma}{81p(p-6)(p-3)\pi^4\calP_\zeta(k_*)} 
\left(\frac{k}{k_*}\right)^{3} 
\ee^{(p-6)\left(N- N_*\right)}\, .
\end{align}
In this case, the integral~(\ref{eq:gnl}) receives its main contribution from its upper bound. Since this bound corresponds to the time at which the trispectrum is evaluated, the
parameter $g_{{}_\mathrm{NL}}^{\mathrm{local}}$ is an explicit function of
time. It increases on super-Hubble scales, contrary to the cases $p<6$ where it reaches a constant value. Here, we have expressed this time dependence in terms of the
number of \efolds~during inflation $N\equiv \ln a$, $N_*$ being $N$ at the time at
which the scale $k_*$ crosses out the Hubble radius.

These analytical approximations are compared with a numerical integration of \Eq{eq:gnl} in \Figs{fig:gNL:p} and~\ref{fig:gNL:k}. One can check that the agreement is excellent (apart from values of $p$ close to but smaller than $p=6$ due to the interpolation mentioned in footnote \ref{footnote:interpolation}). In the left panel of \Fig{fig:gNL:p}, $g_{{}_\mathrm{NL}}$ is displayed as a function of $p$ for a few values of $H\lE$, and one can check that the value of $H\lE$ plays a role only for $p<4$. In the right panel, a few values of $N-N_*$ are shown and one can check that only for $p>6$ does the value of $N-N_*$ play a role (as long as it is positive, \ie on super-Hubble scales).

\section{Observational constraints}
\label{sec:constraints}

\begin{figure}[t]
\begin{center}
\includegraphics[width=0.98\textwidth]{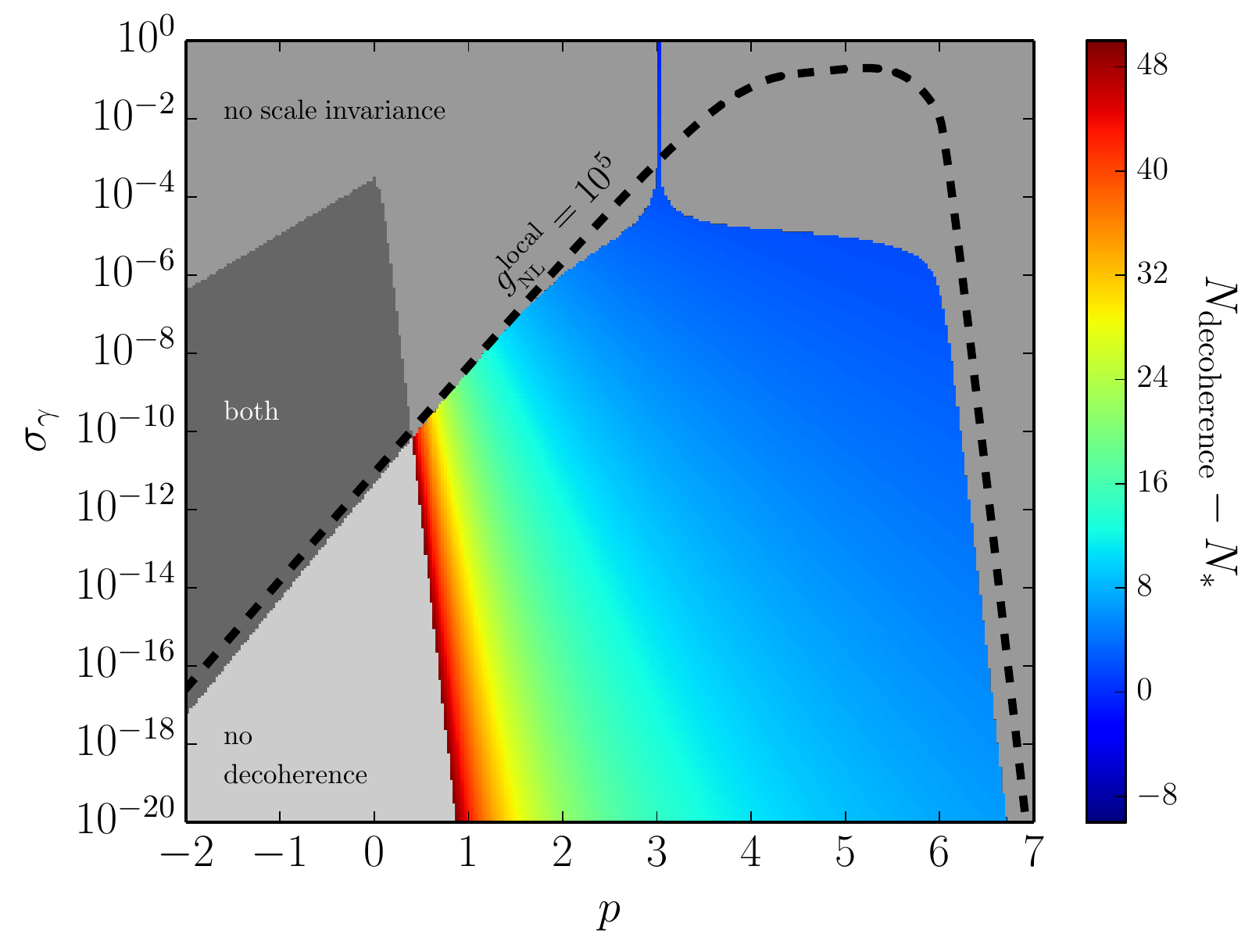}
\caption{Parameter space describing the interaction with the environment ($\sigma_\gamma$ parametrises the interaction strength and $p$ parametrises the rate at which it increases). The region where decoherence does not proceed is denoted ``no decoherence'', and the one where the modifications to the power spectrum are
too large to preserve its quasi scale invariance are denoted ``no scale invariance'' (``both'' referring to where decoherence does not take place \emph{and} quasi scale invariance is spoilt). The constraints coming from non Gaussianities (trispectrum), $\vert g_{{}_\mathrm{NL} \vert}<10^5$,  are such that the region above the black thick dashed line is excluded. This plot corresponds to $H\lE=10^{-3}$, $N-N_*=50$, $\alpha=\pi/4$ and a total number of inflationary \efolds~equal to $10^4$. The colour bar indicates the number of
 \efolds~(with respect to $N_*$, the number of e-folds at which $k_*$
  crosses out the Hubble radius) at which decoherence can be
  considered to be achieved.}
\label{fig:map10m3}
\end{center}
\end{figure}

\begin{figure}[t]
\begin{center}
\includegraphics[width=0.98\textwidth]{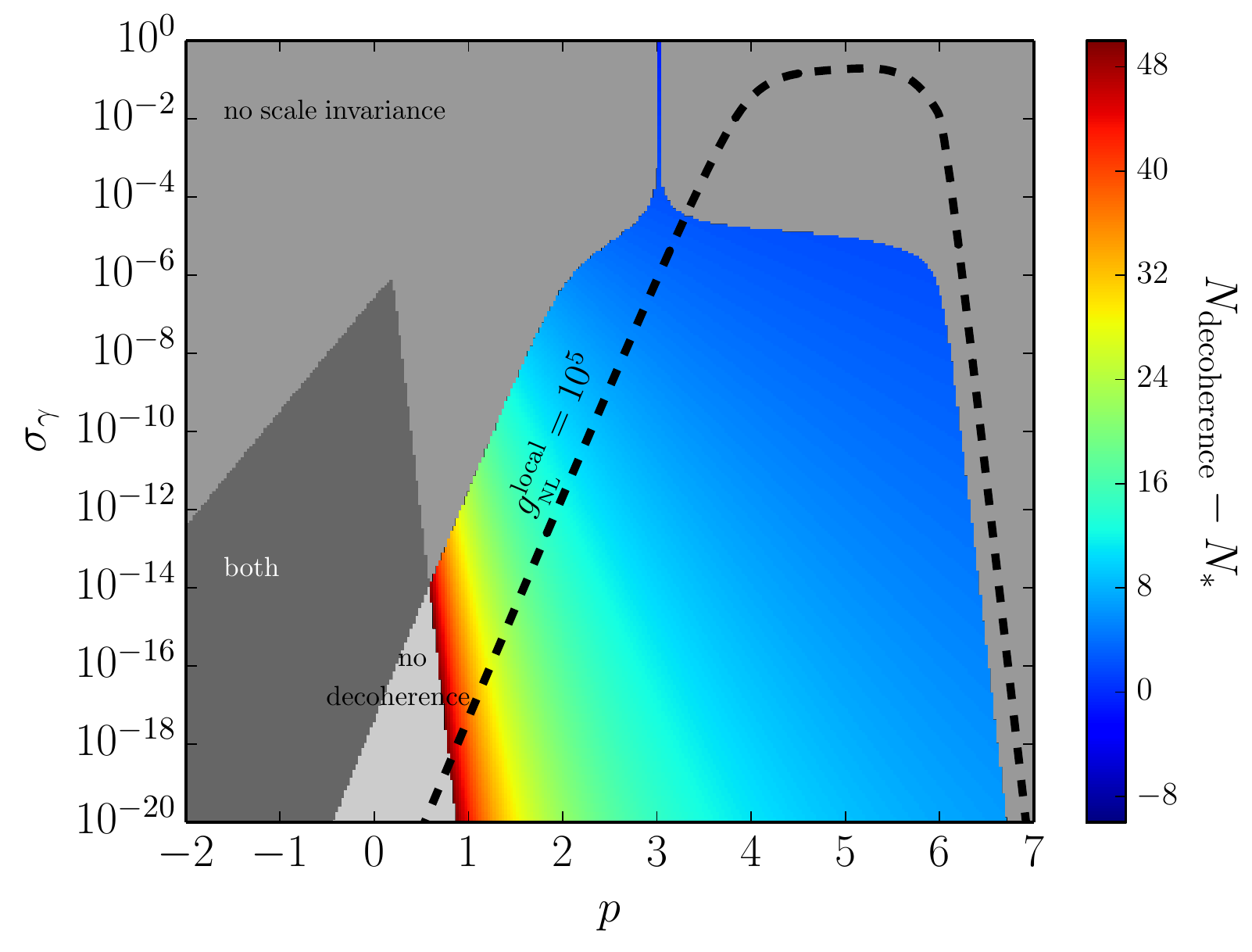}
\caption{Same as \Fig{fig:map10m3} but with
  $H\lE=10^{-6}$.}
\label{fig:map10m6}
\end{center}
\end{figure}

In this section, we use the above results to derive
observational constraints on the interaction strength with the environment, hence on the amount of decoherence cosmological perturbations can have undergone in the early Universe. Concretely, this means constraints on the
parameters $\sigma_\gamma $ and $p$. In \Ref{Martin:2018zbe}, this was
already done using the fact that the power spectrum is also modified
by the interaction between the perturbations and the environment. In
that reference, the constraints were summarized in figures similar to
\Figs{fig:map10m3} and~\ref{fig:map10m6} in the present paper (\Fig{fig:map10m3} corresponds to Fig.~8 of \Ref{Martin:2018zbe} where we have superimposed the constraint from non Gaussianity). In
the space $(p,\sigma_\gamma)$, the grey region labeled ``no scale
invariance'' is a region that is excluded since the modifications
caused by decoherence to the power spectrum spoil the quasi scale invariance to a level which is not observationally acceptable. The
light grey region labeled ``no decoherence'' is a region where the
corrections to the power spectrum are acceptable but where the
strength of the interaction is so small that decoherence itself does
not take place.
% \VV{I suggest removing this sentence, let us be agnostic about decoherence: In that case, the off-diagonal terms of the system density matrix do not decay enough and the system cannot be considered as ``classical'' which is theoretically problematic (although, the question of whether a quantum system is classical is a subtle one, especially in the context of cosmology)}. 
The region ``both'' is a region where both actually happen: the interaction strength is too low to yield decoherence, but already too large to preserve the quasi scale invariance of the power spectrum. The
coloured region is where decoherence occurs without modifying
too much the power spectrum. The colour code, indicated by the colour
bar on the right-hand side of the plot, indicates the number of
\efolds~at which decoherence takes place for the scale $k_*$, measured with respect to a
reference time $N_*$ that corresponds to when the scale $k_*$ crosses out
the Hubble radius during inflation. 

A striking property of these
figures is the vertical blue thin line centred at $p=3$. This
corresponds to a situation where the corrections to the power spectrum
originating from decoherence are themselves almost scale invariant [it is interesting to notice that for $p=3$ we also find $g_{{}_\mathrm{NL}}$ to be scale invariant, see \Eq{eq:gNL:pLT4} and \Fig{fig:gNL:k}]. In
that case, clearly, no constraint on $\sigma_{\gamma }$ can be
set from the power spectrum. Summarising the results of \Ref{Martin:2018zbe}, it
was shown that quasi scale invariance imposes
$\sigma_{\gamma }\ll (H\lE)^{2-p}/N_{_\mathrm{T}}$ if
$p<2$, where $N_{_\mathrm{T}}$ is the total number of \efolds~during
inflation; if $2<p<6$, one has $\sigma_\gamma \ll
1/N_{_\mathrm{T}}$, except if $p=3$ as just mentioned; and for $p>6$, one obtains
$\sigma_\gamma \ll e^{(6-p)(N-N_*)}/N_{_\mathrm{T}}$. Let us notice that the correction to the power spectrum and $g_{{}_\mathrm{NL}}$ have similar properties: for $p<2$, both depend on $H\lE$, for $p>6$, both increase on super-Hubble scales, while for $4<p<6$ both are independent of $H\lE$ and freeze on super-Hubble scales. The only difference is when $2<p<4$, where the corrections to the power spectrum are independent of $H\lE$ and roughly independent of $p$, contrary to the trispectrum.

For this reason, the constraints from the trispectrum are qualitatively similar to the ones from the power spectrum, though quantitatively different. As already mentioned, CMB measurements
indicate that the parameter $g_{{}_\mathrm{NL}}^\mathrm{local}$ is
such that $g_{{}_\mathrm{NL}}^\mathrm{local}=(-9.0+7.7)\times 10^4$ at
the one-sigma level~\cite{Ade:2015ava}. Let us notice that this
constraint only applies to a constant, scale-independent
$g_{{}_\mathrm{NL}}^\mathrm{local}$, which not the
case here since $g_{{}_\mathrm{NL}}^\mathrm{local}$ explicitly depends on $\alpha$ (except if $p>4$) and on $k$ (except if $p=3$, see \Fig{fig:gNL:k}). In principle, one should redo the Planck analysis and
derive new constraints for the specific type of trispectrum
originating from decoherence. Such an analysis is interesting but
clearly beyond the scope of this paper. In addition, this would only
improve the constraints discussed here which can thus be viewed as
conservative.

Using the results of \Sec{subsec:paramgnl}, in  \Figs{fig:map10m3} and~\ref{fig:map10m6} we have superimposed
the constraint coming from the non observation of a trispectrum in
the CMB, see the
black dashed line (the region above this line being
excluded since it leads to $\vert g_{{}_\mathrm{NL}} \vert > 10^{5}$). \Fig{fig:map10m3} corresponds to
$H\lE=10^{-3}$ while \Fig{fig:map10m6} is for
$H\lE=10^{-6}$. Technically, evaluating the expressions of
$g_{{}_\mathrm{NL}}^\mathrm{local}$ derived in
\Sec{subsec:paramgnl} at $k=k_*$, one finds
\bea
\sigma_\gamma <
\begin{cases}
\dfrac{81}{25}(4-p)2^{8-p}\pi^4 \calP_\zeta\left(k_*\right) \sigma\left(g_{{}_\mathrm{NL}}^{\mathrm{local}}\right) (H \lE)^{4-p} \simeq \mathcal{O}\left(10^{-2}\right)(H \lE)^{4-p}
\quad\mathrm{if}\quad p<4\, ,
\vspace{0.4cm}\\
\dfrac{243}{25}\pi^4  \calP_\zeta\left(k_*\right)
\sigma\left(g_{{}_\mathrm{NL}}^{\mathrm{local}}\right) \simeq 0.2
\quad\mathrm{if}\quad 4<p<6\, ,
\vspace{0.4cm}\\
\dfrac{81}{50} \pi^4 p(p-6)(p-3)
\sigma\left(g_{{}_\mathrm{NL}}^{\mathrm{local}}\right)
\calP_\zeta\left(k_*\right) \ee^{(6-p)\left(N-N_*\right)} \simeq
%\mathcal{O}(1)
\ee^{(6-p)\left(N-N_*\right)}
\quad\mathrm{if}\quad p>6\, ,
\end{cases}\
\eea
%$\sigma_\gamma\ll 81(4-p)2^{8-p}\pi^4/25 \times \calP_\zeta\left(k_*\right) \sigma\left(g_{{}_\mathrm{NL}}^{\mathrm{local}}\right) (H \lE)^{4-p} \simeq \mathcal{O}\left(10^{-2}\right)(H \lE)^{4-p}$, if $p<4$. If $4<p<6$ then $\sigma_\gamma\ll 243\pi^4/25 \times \calP_\zeta\left(k_*\right) \sigma\left(g_{{}_\mathrm{NL}}^{\mathrm{local}}\right) \simeq 0.2$. Finally, if $p>6$, one obtains $\sigma_\gamma\ll 81 \pi^4 p(p-6)(p-3)/50 \times\sigma\left(g_{{}_\mathrm{NL}}^{\mathrm{local}}\right) \calP_\zeta\left(k_*\right) \ee^{(6-p)\Delta N_*} \simeq \mathcal{O}(1)\ee^{(6-p)\Delta N_*}$.
where $\sigma\left(g_{{}_\mathrm{NL}}^{\mathrm{local}}\right)\sim 10^5$ denotes the one-sigma contraint on $\vert g_{{}_\mathrm{NL}}^{\mathrm{local}} \vert$.
From these expressions and from \Figs{fig:map10m3}
and~\ref{fig:map10m6}, one sees that when $p<4$, the most stringent
constraint comes from the trispectrum while when $p>4$, it still comes
from the power spectrum. To our knowledge, this is a rare example in
cosmology where the constraints coming from the trispectrum can be more
efficient than those coming from the power spectrum. 

An important conclusion is the fact that, for $p=3$ (which in \Ref{Martin:2018zbe} was shown to correspond to the case where the environment is made of a heavy test scalar fied), the vertical blue
line discussed before is now excluded, if $H\lE$ is sufficiently small (for $H\lE\simeq 10^{-3}$ there is still a small portion of the line that is viable). In \Ref{Martin:2018zbe}, it was shown that along the blue line, the corrections to the power spectrum are quasi scale-invariant but can still substantially change the predictions of a given model of inflation, thus opening up the possibility to detect decoherence effects in the power spectrum. A non-trivial consequence of the current work is that, because of the constraints on the trispectrum, this cannot happen for sufficiently small $H\lE$, since current bounds on non Gaussianities already exclude such a possibility.

\section{Conclusions}
\label{sec:conclusion}

Let us now recap our main findings. In this article, we have argued
that cosmological perturbations should be treated as an open quantum system rather than a close system.
In that case, decoherence can take place,
which may be important when one tries to understand the quantum-to-classical transition of cosmological perturbations, or if one wonders whether genuine quantum correlations can be observed in cosmological perturbations that would unveil their quantum mechanical origin~\cite{Martin:2015qta, Martin:2017zxs}.

The interaction of the system with an
environment also induces modifications to observable quantities such as the power spectrum~\cite{Martin:2018zbe}. In this work we have shown that non Gaussianities also unavoidably appear. These non
Gaussianities show up at the level of the four-point correlation
function (trispectrum) while the bispectrum still vanishes. Using the Planck constraint on $g_{{}_\mathrm{NL}}^{\mathrm{local}}$,
we have derived new constraints on the strength of the interaction
between the cosmological perturbations and the
environment. Remarkably, in some regimes, these constraints are more
stringent than those inferred from the two-point correlation function.

Despite the generic character of these results, some important
questions remain to be addressed. As mentioned before, the
Planck constraints on non Gaussianities we have used were derived assuming a scale-invariant
$g_{{}_\mathrm{NL}}^{\mathrm{local}}$ parameter, while the
$g_{{}_\mathrm{NL}}^{\mathrm{local}}$ parameter caused by decoherence is
generally scale dependent.
Let us also recall that we have focused on the trispectrum in the equilateral configuration, since in that case it can be obtained from a  differential equation of order five that we could solve analytically. For an arbitrary configuration, one has to solve a  differential equation of order sixteen, which makes the analysis more complicated but still numerically achievable. Although our analysis leads to conservative
bounds, it would be interesting to improve them by testing the actual
trispectrum pattern found in this paper.

Another
interesting question is the form of the interaction assumed between
the perturbations and the environment. Here, we have considered the
cases of linear (for which there are no non Gaussianities at leading order in the interaction strength) and quadratic (for which non Gaussianities show up only in the trispectrum) interactions. Clearly, it would be
important to consider higher-order interactions (for which we expect non-Gaussianities in higher correlation functions, since an interaction of order $n$ gives rise to non-vanishing connected $m$-point correlation functions for all $m$ even and smaller than $2n$). This was
done for the power spectrum in \Ref{Martin:2018zbe} where a
diagrammatic calculation of the source $S_n$ is
available. Unfortunately, such a tool does not seem to be available
for non Gaussianities which makes the problem much harder. We hope to
come back to these issues in the future.
\begin{acknowledgments}
  V.~V. acknowledges funding from the European Union's Horizon 2020
  research and innovation programme under the Marie Sk\l odowska-Curie
  grant agreement N${}^0$ 750491.
\end{acknowledgments}
\appendix

\section{Correlators for linear interactions}
\label{sec:applinear}

As already discussed at the beginning of \Sec{sec:highorder}, the case of a linear interaction is
peculiar since it has been shown in \Ref{Martin:2018zbe} that
the density matrix of cosmological perturbations remains
Gaussian. Therefore, there is no need to calculate the higher-order
correlation functions since the result is entirely given by the Wick
theorem. As a consistency check, it is nevertheless
interesting to perform the calculation of the three- and four-point correlation
functions using \Eq{eq:Lindblad:mean:Fourrier} in order to see explicitly how Gaussianity is preserved.

The equations for the two-point correlation functions were established
in \Ref{Martin:2018zbe} and read
\begin{align}
\label{eq:vvlinear}
\frac{\dd  }{\dd  \eta} \left\langle \hat{v}_{{\bm k}_1}
\hat{v}_{{\bm k}_2}\right\rangle &=\left \langle 
\hat{v}_{{\bm k}_1}
\hat{p}_{{\bm k}_2}
+\hat{p}_{{\bm k}_1}
\hat{v}_{{\bm k}_2}
\right\rangle, \\
\label{eq:vplinear}
\frac{\dd  }{\dd  \eta} \left\langle \hat{v}_{{\bm k}_1}
\hat{p}_{{\bm k}_2}\right\rangle &=\left \langle 
\hat{p}_{{\bm k}_1}
\hat{p}_{{\bm k}_2}\right \rangle
-\omega^2(k_2)\left\langle
\hat{v}_{{\bm k}_1}
\hat{v}_{{\bm k}_2}
\right\rangle,\\
\label{eq:pvlinear}
\frac{\dd  }{\dd  \eta} \left\langle \hat{p}_{{\bm k}_1}
\hat{v}_{{\bm k}_2}\right\rangle &=\left \langle 
\hat{p}_{{\bm k}_1}
\hat{p}_{{\bm k}_2}\right \rangle
-\omega^2(k_1)\left\langle
\hat{v}_{{\bm k}_1}
\hat{v}_{{\bm k}_2}
\right\rangle,\\
\label{eq:pplinear}
\frac{\dd  }{\dd  \eta} \left\langle \hat{p}_{{\bm k}_1}
\hat{p}_{{\bm k}_2}\right\rangle &=
-\omega^2(k_2)\left\langle\hat{p}_{{\bm k}_1}
\hat{v}_{{\bm k}_2}\right \rangle
-\omega^2(k_1)\left\langle
\hat{v}_{{\bm k}_1}
\hat{p}_{{\bm k}_2}
\right\rangle
+\gamma (2\pi)^{3/2}
\tilde{C}_{R}({\bm k}_1)\delta ({\bm k}_1+{\bm k}_2).
\end{align}
The presence of the environment manifests itself by the term
proportional to $\gamma$ only in the evolution equation for
$\left\langle \hat{p}_{\bm{k}_1}\hat{p}_{{\bm k}_2}\right\rangle$. It
is responsible for the appearance of a modified power spectrum.

The calculation of the three-point correlators proceeds in the same 
way and one obtains
\begin{align}
\label{eq:3points}
\frac{\dd  }{\dd  \eta} \left\langle \hat{v}_{{\bm k}_1}
\hat{v}_{{\bm k}_2}\hat{v}_{{\bm k}_3}\right\rangle &=
\left \langle 
\hat{v}_{{\bm k}_1}
\hat{v}_{{\bm k}_2}
\hat{p}_{{\bm k}_3}\right\rangle
+\left \langle 
\hat{v}_{{\bm k}_1}
\hat{p}_{{\bm k}_2}
\hat{v}_{{\bm k}_3}\right\rangle
+
\left \langle 
\hat{p}_{{\bm k}_1}
\hat{v}_{{\bm k}_2}
\hat{v}_{{\bm k}_3}\right\rangle, \\
\frac{\dd  }{\dd  \eta} \left\langle \hat{v}_{{\bm k}_1}
\hat{v}_{{\bm k}_2}\hat{p}_{{\bm k}_3}\right\rangle &=
\left \langle 
\hat{v}_{{\bm k}_1}
\hat{p}_{{\bm k}_2}
\hat{p}_{{\bm k}_3}\right\rangle
+\left \langle 
\hat{p}_{{\bm k}_1}
\hat{v}_{{\bm k}_2}
\hat{p}_{{\bm k}_3}\right\rangle
-\omega^2(k_3)
\left \langle 
\hat{v}_{{\bm k}_1}
\hat{v}_{{\bm k}_2}
\hat{v}_{{\bm k}_3}\right\rangle, \\
\frac{\dd  }{\dd  \eta} \left\langle \hat{v}_{{\bm k}_1}
\hat{p}_{{\bm k}_2}\hat{v}_{{\bm k}_3}\right\rangle &=
\left \langle 
\hat{v}_{{\bm k}_1}
\hat{p}_{{\bm k}_2}
\hat{p}_{{\bm k}_3}\right\rangle
+\left \langle 
\hat{p}_{{\bm k}_1}
\hat{p}_{{\bm k}_2}
\hat{v}_{{\bm k}_3}\right\rangle
-\omega^2(k_2)
\left \langle 
\hat{v}_{{\bm k}_1}
\hat{v}_{{\bm k}_2}
\hat{v}_{{\bm k}_3}\right\rangle, \\
\frac{\dd  }{\dd  \eta} \left\langle \hat{p}_{{\bm k}_1}
\hat{v}_{{\bm k}_2}\hat{v}_{{\bm k}_3}\right\rangle &=
\left \langle 
\hat{p}_{{\bm k}_1}
\hat{v}_{{\bm k}_2}
\hat{p}_{{\bm k}_3}\right\rangle
+\left \langle 
\hat{p}_{{\bm k}_1}
\hat{p}_{{\bm k}_2}
\hat{v}_{{\bm k}_3}\right\rangle
-\omega^2(k_1)
\left \langle 
\hat{v}_{{\bm k}_1}
\hat{v}_{{\bm k}_2}
\hat{v}_{{\bm k}_3}\right\rangle, \\
\frac{\dd  }{\dd  \eta} \left\langle \hat{v}_{{\bm k}_1}
\hat{p}_{{\bm k}_2}\hat{p}_{{\bm k}_3}\right\rangle &=
\left \langle 
\hat{p}_{{\bm k}_1}
\hat{p}_{{\bm k}_2}
\hat{p}_{{\bm k}_3}\right\rangle
-\omega^2(k_3)
\left \langle 
\hat{v}_{{\bm k}_1}
\hat{p}_{{\bm k}_2}
\hat{v}_{{\bm k}_3}\right\rangle
-\omega^2(k_2)
\left \langle 
\hat{v}_{{\bm k}_1}
\hat{v}_{{\bm k}_2}
\hat{p}_{{\bm k}_3}\right\rangle
, \\
\frac{\dd  }{\dd  \eta} \left\langle \hat{p}_{{\bm k}_1}
\hat{v}_{{\bm k}_2}\hat{p}_{{\bm k}_3}\right\rangle &=
\left \langle 
\hat{p}_{{\bm k}_1}
\hat{p}_{{\bm k}_2}
\hat{p}_{{\bm k}_3}\right\rangle
-\omega^2(k_1)
\left \langle 
\hat{v}_{{\bm k}_1}
\hat{v}_{{\bm k}_2}
\hat{p}_{{\bm k}_3}\right\rangle
-\omega^2(k_3)
\left \langle 
\hat{p}_{{\bm k}_1}
\hat{v}_{{\bm k}_2}
\hat{v}_{{\bm k}_3}\right\rangle
, \\
\frac{\dd  }{\dd  \eta} \left\langle \hat{p}_{{\bm k}_1}
\hat{p}_{{\bm k}_2}\hat{v}_{{\bm k}_3}\right\rangle &=
\left \langle 
\hat{p}_{{\bm k}_1}
\hat{p}_{{\bm k}_2}
\hat{p}_{{\bm k}_3}\right\rangle
-\omega^2(k_2)
\left \langle 
\hat{p}_{{\bm k}_1}
\hat{v}_{{\bm k}_2}
\hat{v}_{{\bm k}_3}\right\rangle
-\omega^2(k_1)
\left \langle 
\hat{v}_{{\bm k}_1}
\hat{p}_{{\bm k}_2}
\hat{v}_{{\bm k}_3}\right\rangle
, \\
\frac{\dd  }{\dd  \eta} \left\langle \hat{p}_{{\bm k}_1}
\hat{p}_{{\bm k}_2}\hat{v}_{{\bm k}_3}\right\rangle &=
-\omega^2(k_1)
\left \langle 
\hat{v}_{{\bm k}_1}
\hat{p}_{{\bm k}_2}
\hat{p}_{{\bm k}_3}\right\rangle
-\omega^2(k_2)
\left \langle 
\hat{p}_{{\bm k}_1}
\hat{v}_{{\bm k}_2}
\hat{p}_{{\bm k}_3}\right\rangle
-\omega^2(k_3)
\left \langle 
\hat{p}_{{\bm k}_1}
\hat{p}_{{\bm k}_2}
\hat{v}_{{\bm k}_3}\right\rangle.
\end{align}
We notice that no term proportional to $\gamma $ is present in the
above equations and this clearly implies that the three-point
correlation functions all vanish if they are initially set to zero, in accordance with the fact that the
system remains Gaussian if it is initially placed in the (Gaussian) Bunch-Davies vacuum state.

Let us now derive the evolution equations of the four-point
correlators. Lengthy but straightforward calculations lead to
\begin{align}
\label{eq:4points}
\frac{\dd  }{\dd  \eta} \left\langle \hat{v}_{{\bm k}_1}
\hat{v}_{{\bm k}_2}\hat{v}_{{\bm k}_3}\hat{v}_{{\bm k}_4}\right\rangle &=
\left \langle 
\hat{v}_{{\bm k}_1}
\hat{v}_{{\bm k}_2}
\hat{v}_{{\bm k}_3}
\hat{p}_{{\bm k}_4}
\right\rangle
+
\left \langle 
\hat{v}_{{\bm k}_1}
\hat{v}_{{\bm k}_2}
\hat{p}_{{\bm k}_3}
\hat{v}_{{\bm k}_4}
\right\rangle
+
\left \langle 
\hat{v}_{{\bm k}_1}
\hat{p}_{{\bm k}_2}
\hat{v}_{{\bm k}_3}
\hat{v}_{{\bm k}_4}
\right\rangle
+
\left \langle 
\hat{p}_{{\bm k}_1}
\hat{v}_{{\bm k}_2}
\hat{v}_{{\bm k}_3}
\hat{v}_{{\bm k}_4}
\right\rangle, 
\\
\frac{\dd  }{\dd  \eta} \left\langle \hat{v}_{{\bm k}_1}
\hat{v}_{{\bm k}_2}\hat{v}_{{\bm k}_3}\hat{p}_{{\bm k}_4}\right\rangle &=
\left \langle 
\hat{v}_{{\bm k}_1}
\hat{v}_{{\bm k}_2}
\hat{p}_{{\bm k}_3}
\hat{p}_{{\bm k}_4}
\right\rangle
+
\left \langle 
\hat{v}_{{\bm k}_1}
\hat{p}_{{\bm k}_2}
\hat{v}_{{\bm k}_3}
\hat{p}_{{\bm k}_4}
\right\rangle
+
\left \langle 
\hat{p}_{{\bm k}_1}
\hat{v}_{{\bm k}_2}
\hat{v}_{{\bm k}_3}
\hat{p}_{{\bm k}_4}
\right\rangle
\nonumber \\
& -\omega^2(k_4)
\left \langle 
\hat{v}_{{\bm k}_1}
\hat{v}_{{\bm k}_2}
\hat{v}_{{\bm k}_3}
\hat{v}_{{\bm k}_4}
\right\rangle, 
\\
\frac{\dd  }{\dd  \eta} \left\langle \hat{v}_{{\bm k}_1}
\hat{v}_{{\bm k}_2}\hat{p}_{{\bm k}_3}\hat{v}_{{\bm k}_4}\right\rangle &=
\left \langle 
\hat{p}_{{\bm k}_1}
\hat{v}_{{\bm k}_2}
\hat{p}_{{\bm k}_3}
\hat{v}_{{\bm k}_4}
\right\rangle
+
\left \langle 
\hat{v}_{{\bm k}_1}
\hat{v}_{{\bm k}_2}
\hat{p}_{{\bm k}_3}
\hat{p}_{{\bm k}_4}
\right\rangle
+
\left \langle 
\hat{v}_{{\bm k}_1}
\hat{p}_{{\bm k}_2}
\hat{p}_{{\bm k}_3}
\hat{v}_{{\bm k}_4}
\right\rangle
\nonumber \\
& -\omega^2(k_3)
\left \langle 
\hat{v}_{{\bm k}_1}
\hat{v}_{{\bm k}_2}
\hat{v}_{{\bm k}_3}
\hat{v}_{{\bm k}_4}
\right\rangle, 
\\
\frac{\dd  }{\dd  \eta} \left\langle \hat{v}_{{\bm k}_1}
\hat{p}_{{\bm k}_2}\hat{v}_{{\bm k}_3}\hat{v}_{{\bm k}_4}\right\rangle &=
\left \langle 
\hat{v}_{{\bm k}_1}
\hat{p}_{{\bm k}_2}
\hat{v}_{{\bm k}_3}
\hat{p}_{{\bm k}_4}
\right\rangle
+
\left \langle 
\hat{v}_{{\bm k}_1}
\hat{p}_{{\bm k}_2}
\hat{p}_{{\bm k}_3}
\hat{v}_{{\bm k}_4}
\right\rangle
+
\left \langle 
\hat{p}_{{\bm k}_1}
\hat{p}_{{\bm k}_2}
\hat{v}_{{\bm k}_3}
\hat{v}_{{\bm k}_4}
\right\rangle
\nonumber \\
& -\omega^2(k_2)
\left \langle 
\hat{v}_{{\bm k}_1}
\hat{v}_{{\bm k}_2}
\hat{v}_{{\bm k}_3}
\hat{v}_{{\bm k}_4}
\right\rangle, 
\\
\frac{\dd  }{\dd  \eta} \left\langle \hat{p}_{{\bm k}_1}
\hat{v}_{{\bm k}_2}\hat{v}_{{\bm k}_3}\hat{v}_{{\bm k}_4}\right\rangle &=
\left \langle 
\hat{p}_{{\bm k}_1}
\hat{v}_{{\bm k}_2}
\hat{v}_{{\bm k}_3}
\hat{p}_{{\bm k}_4}
\right\rangle
+
\left \langle 
\hat{p}_{{\bm k}_1}
\hat{v}_{{\bm k}_2}
\hat{p}_{{\bm k}_3}
\hat{v}_{{\bm k}_4}
\right\rangle
+
\left \langle 
\hat{p}_{{\bm k}_1}
\hat{p}_{{\bm k}_2}
\hat{v}_{{\bm k}_3}
\hat{v}_{{\bm k}_4}
\right\rangle
\nonumber \\
& -\omega^2(k_1)
\left \langle 
\hat{v}_{{\bm k}_1}
\hat{v}_{{\bm k}_2}
\hat{v}_{{\bm k}_3}
\hat{v}_{{\bm k}_4}
\right\rangle, 
\\
\frac{\dd  }{\dd  \eta} \left\langle \hat{v}_{{\bm k}_1}
\hat{v}_{{\bm k}_2}\hat{p}_{{\bm k}_3}\hat{p}_{{\bm k}_4}\right\rangle &=
\left \langle 
\hat{v}_{{\bm k}_1}
\hat{p}_{{\bm k}_2}
\hat{p}_{{\bm k}_3}
\hat{p}_{{\bm k}_4}
\right\rangle
+
\left \langle 
\hat{p}_{{\bm k}_1}
\hat{v}_{{\bm k}_2}
\hat{p}_{{\bm k}_3}
\hat{p}_{{\bm k}_4}
\right\rangle
-\omega^2(k_4)
\left \langle 
\hat{v}_{{\bm k}_1}
\hat{v}_{{\bm k}_2}
\hat{p}_{{\bm k}_3}
\hat{v}_{{\bm k}_4}
\right\rangle
\nonumber \\
& -\omega^2(k_3)
\left \langle 
\hat{v}_{{\bm k}_1}
\hat{v}_{{\bm k}_2}
\hat{v}_{{\bm k}_3}
\hat{p}_{{\bm k}_4}
\right\rangle
+\gamma(2\pi)^{3/2}\tilde{C}_R(k_3)
\left \langle 
\hat{v}_{{\bm k}_1}
\hat{v}_{{\bm k}_2}
\right\rangle
\delta \left({\bm k}_3+{\bm k}_4\right), 
\\
\frac{\dd  }{\dd  \eta} \left\langle \hat{v}_{{\bm k}_1}
\hat{p}_{{\bm k}_2}\hat{v}_{{\bm k}_3}\hat{p}_{{\bm k}_4}\right\rangle &=
\left \langle 
\hat{v}_{{\bm k}_1}
\hat{p}_{{\bm k}_2}
\hat{p}_{{\bm k}_3}
\hat{p}_{{\bm k}_4}
\right\rangle
+
\left \langle 
\hat{p}_{{\bm k}_1}
\hat{p}_{{\bm k}_2}
\hat{v}_{{\bm k}_3}
\hat{p}_{{\bm k}_4}
\right\rangle
-\omega^2(k_4)
\left \langle 
\hat{v}_{{\bm k}_1}
\hat{p}_{{\bm k}_2}
\hat{v}_{{\bm k}_3}
\hat{v}_{{\bm k}_4}
\right\rangle
\nonumber \\
& -\omega^2(k_2)
\left \langle 
\hat{v}_{{\bm k}_1}
\hat{v}_{{\bm k}_2}
\hat{v}_{{\bm k}_3}
\hat{p}_{{\bm k}_4}
\right\rangle
+\gamma(2\pi)^{3/2}\tilde{C}_R(k_2)
\left \langle 
\hat{v}_{{\bm k}_1}
\hat{v}_{{\bm k}_3}
\right\rangle
\delta \left({\bm k}_2+{\bm k}_4\right), 
\\
\frac{\dd  }{\dd  \eta} \left\langle \hat{p}_{{\bm k}_1}
\hat{v}_{{\bm k}_2}\hat{v}_{{\bm k}_3}\hat{p}_{{\bm k}_4}\right\rangle &=
\left \langle 
\hat{p}_{{\bm k}_1}
\hat{v}_{{\bm k}_2}
\hat{p}_{{\bm k}_3}
\hat{p}_{{\bm k}_4}
\right\rangle
+
\left \langle 
\hat{p}_{{\bm k}_1}
\hat{p}_{{\bm k}_2}
\hat{v}_{{\bm k}_3}
\hat{p}_{{\bm k}_4}
\right\rangle
-\omega^2(k_4)
\left \langle 
\hat{p}_{{\bm k}_1}
\hat{v}_{{\bm k}_2}
\hat{v}_{{\bm k}_3}
\hat{v}_{{\bm k}_4}
\right\rangle
\nonumber \\
& -\omega^2(k_1)
\left \langle 
\hat{v}_{{\bm k}_1}
\hat{v}_{{\bm k}_2}
\hat{v}_{{\bm k}_3}
\hat{p}_{{\bm k}_4}
\right\rangle
+\gamma(2\pi)^{3/2}\tilde{C}_R(k_1)
\left \langle 
\hat{v}_{{\bm k}_2}
\hat{v}_{{\bm k}_3}
\right\rangle
\delta \left({\bm k}_1+{\bm k}_4\right),
\\
\frac{\dd  }{\dd  \eta} \left\langle \hat{v}_{{\bm k}_1}
\hat{p}_{{\bm k}_2}\hat{p}_{{\bm k}_3}\hat{v}_{{\bm k}_4}\right\rangle &=
\left \langle 
\hat{v}_{{\bm k}_1}
\hat{p}_{{\bm k}_2}
\hat{p}_{{\bm k}_3}
\hat{p}_{{\bm k}_4}
\right\rangle
+
\left \langle 
\hat{p}_{{\bm k}_1}
\hat{p}_{{\bm k}_2}
\hat{p}_{{\bm k}_3}
\hat{v}_{{\bm k}_4}
\right\rangle
-\omega^2(k_3)
\left \langle 
\hat{v}_{{\bm k}_1}
\hat{p}_{{\bm k}_2}
\hat{v}_{{\bm k}_3}
\hat{v}_{{\bm k}_4}
\right\rangle
\nonumber \\
& -\omega^2(k_2)
\left \langle 
\hat{v}_{{\bm k}_1}
\hat{v}_{{\bm k}_2}
\hat{p}_{{\bm k}_3}
\hat{v}_{{\bm k}_4}
\right\rangle
+\gamma(2\pi)^{3/2}\tilde{C}_R(k_2)
\left \langle 
\hat{v}_{{\bm k}_1}
\hat{v}_{{\bm k}_4}
\right\rangle
\delta \left({\bm k}_2+{\bm k}_3\right),
\\
\frac{\dd  }{\dd  \eta} \left\langle \hat{p}_{{\bm k}_1}
\hat{v}_{{\bm k}_2}\hat{p}_{{\bm k}_3}\hat{v}_{{\bm k}_4}\right\rangle &=
\left \langle 
\hat{p}_{{\bm k}_1}
\hat{v}_{{\bm k}_2}
\hat{p}_{{\bm k}_3}
\hat{p}_{{\bm k}_4}
\right\rangle
+
\left \langle 
\hat{p}_{{\bm k}_1}
\hat{p}_{{\bm k}_2}
\hat{p}_{{\bm k}_3}
\hat{v}_{{\bm k}_4}
\right\rangle
-\omega^2(k_3)
\left \langle 
\hat{p}_{{\bm k}_1}
\hat{v}_{{\bm k}_2}
\hat{v}_{{\bm k}_3}
\hat{v}_{{\bm k}_4}
\right\rangle
\nonumber \\
& -\omega^2(k_1)
\left \langle 
\hat{v}_{{\bm k}_1}
\hat{v}_{{\bm k}_2}
\hat{p}_{{\bm k}_3}
\hat{v}_{{\bm k}_4}
\right\rangle
+\gamma(2\pi)^{3/2}\tilde{C}_R(k_1)
\left \langle 
\hat{v}_{{\bm k}_2}
\hat{v}_{{\bm k}_4}
\right\rangle
\delta \left({\bm k}_1+{\bm k}_3\right),
\\
\frac{\dd  }{\dd  \eta} \left\langle \hat{p}_{{\bm k}_1}
\hat{p}_{{\bm k}_2}\hat{v}_{{\bm k}_3}\hat{v}_{{\bm k}_4}\right\rangle &=
\left \langle 
\hat{p}_{{\bm k}_1}
\hat{p}_{{\bm k}_2}
\hat{v}_{{\bm k}_3}
\hat{p}_{{\bm k}_4}
\right\rangle
+
\left \langle 
\hat{p}_{{\bm k}_1}
\hat{p}_{{\bm k}_2}
\hat{p}_{{\bm k}_3}
\hat{v}_{{\bm k}_4}
\right\rangle
-\omega^2(k_2)
\left \langle 
\hat{p}_{{\bm k}_1}
\hat{v}_{{\bm k}_2}
\hat{v}_{{\bm k}_3}
\hat{v}_{{\bm k}_4}
\right\rangle
\nonumber \\
& -\omega^2(k_1)
\left \langle 
\hat{v}_{{\bm k}_1}
\hat{p}_{{\bm k}_2}
\hat{v}_{{\bm k}_3}
\hat{v}_{{\bm k}_4}
\right\rangle
+\gamma(2\pi)^{3/2}\tilde{C}_R(k_1)
\left \langle 
\hat{v}_{{\bm k}_3}
\hat{v}_{{\bm k}_4}
\right\rangle
\delta \left({\bm k}_1+{\bm k}_2\right),
\\
\frac{\dd  }{\dd  \eta} \left\langle \hat{v}_{{\bm k}_1}
\hat{p}_{{\bm k}_2}\hat{p}_{{\bm k}_3}\hat{p}_{{\bm k}_4}\right\rangle &=
\left \langle 
\hat{p}_{{\bm k}_1}
\hat{p}_{{\bm k}_2}
\hat{p}_{{\bm k}_3}
\hat{p}_{{\bm k}_4}
\right\rangle
-\omega^2(k_4)
\left \langle 
\hat{v}_{{\bm k}_1}
\hat{p}_{{\bm k}_2}
\hat{p}_{{\bm k}_3}
\hat{v}_{{\bm k}_4}
\right\rangle
-\omega^2(k_3)
\left \langle 
\hat{v}_{{\bm k}_1}
\hat{p}_{{\bm k}_2}
\hat{v}_{{\bm k}_3}
\hat{p}_{{\bm k}_4}
\right\rangle
\nonumber \\
& -\omega^2(k_2)
\left \langle 
\hat{v}_{{\bm k}_1}
\hat{v}_{{\bm k}_2}
\hat{p}_{{\bm k}_3}
\hat{p}_{{\bm k}_4}
\right\rangle
+\gamma(2\pi)^{3/2}\tilde{C}_R(k_3)
\left \langle 
\hat{v}_{{\bm k}_1}
\hat{p}_{{\bm k}_2}
\right\rangle
\delta \left({\bm k}_3+{\bm k}_4\right)
\nonumber \\ &
+\gamma(2\pi)^{3/2}\tilde{C}_R(k_2)
\left \langle 
\hat{v}_{{\bm k}_1}
\hat{p}_{{\bm k}_3}
\right\rangle
\delta \left({\bm k}_2+{\bm k}_4\right)
\nonumber \\ &
+\gamma(2\pi)^{3/2}\tilde{C}_R(k_2)
\left \langle 
\hat{v}_{{\bm k}_1}
\hat{p}_{{\bm k}_4}
\right\rangle
\delta \left({\bm k}_2+{\bm k}_3\right), 
\\
\frac{\dd  }{\dd  \eta} \left\langle \hat{p}_{{\bm k}_1}
\hat{v}_{{\bm k}_2}\hat{p}_{{\bm k}_3}\hat{p}_{{\bm k}_4}\right\rangle &=
\left \langle 
\hat{p}_{{\bm k}_1}
\hat{p}_{{\bm k}_2}
\hat{p}_{{\bm k}_3}
\hat{p}_{{\bm k}_4}
\right\rangle
-\omega^2(k_4)
\left \langle 
\hat{p}_{{\bm k}_1}
\hat{v}_{{\bm k}_2}
\hat{p}_{{\bm k}_3}
\hat{v}_{{\bm k}_4}
\right\rangle
-\omega^2(k_3)
\left \langle 
\hat{p}_{{\bm k}_1}
\hat{v}_{{\bm k}_2}
\hat{v}_{{\bm k}_3}
\hat{p}_{{\bm k}_4}
\right\rangle
\nonumber \\
& -\omega^2(k_1)
\left \langle 
\hat{v}_{{\bm k}_1}
\hat{v}_{{\bm k}_2}
\hat{p}_{{\bm k}_3}
\hat{p}_{{\bm k}_4}
\right\rangle
+\gamma(2\pi)^{3/2}\tilde{C}_R(k_3)
\left \langle 
\hat{p}_{{\bm k}_1}
\hat{v}_{{\bm k}_2}
\right\rangle
\delta \left({\bm k}_3+{\bm k}_4\right)
\nonumber \\ &
+\gamma(2\pi)^{3/2}\tilde{C}_R(k_1)
\left \langle 
\hat{v}_{{\bm k}_2}
\hat{p}_{{\bm k}_3}
\right\rangle
\delta \left({\bm k}_1+{\bm k}_4\right)
\nonumber \\ &
+\gamma(2\pi)^{3/2}\tilde{C}_R(k_1)
\left \langle 
\hat{v}_{{\bm k}_2}
\hat{p}_{{\bm k}_4}
\right\rangle
\delta \left({\bm k}_1+{\bm k}_3\right), 
\\
\frac{\dd  }{\dd  \eta} \left\langle \hat{p}_{{\bm k}_1}
\hat{p}_{{\bm k}_2}\hat{v}_{{\bm k}_3}\hat{p}_{{\bm k}_4}\right\rangle &=
\left \langle 
\hat{p}_{{\bm k}_1}
\hat{p}_{{\bm k}_2}
\hat{p}_{{\bm k}_3}
\hat{p}_{{\bm k}_4}
\right\rangle
-\omega^2(k_4)
\left \langle 
\hat{p}_{{\bm k}_1}
\hat{p}_{{\bm k}_2}
\hat{v}_{{\bm k}_3}
\hat{v}_{{\bm k}_4}
\right\rangle
-\omega^2(k_2)
\left \langle 
\hat{p}_{{\bm k}_1}
\hat{v}_{{\bm k}_2}
\hat{v}_{{\bm k}_3}
\hat{p}_{{\bm k}_4}
\right\rangle
\nonumber \\
& -\omega^2(k_1)
\left \langle 
\hat{v}_{{\bm k}_1}
\hat{p}_{{\bm k}_2}
\hat{v}_{{\bm k}_3}
\hat{p}_{{\bm k}_4}
\right\rangle
+\gamma(2\pi)^{3/2}\tilde{C}_R(k_2)
\left \langle 
\hat{p}_{{\bm k}_1}
\hat{v}_{{\bm k}_3}
\right\rangle
\delta \left({\bm k}_2+{\bm k}_4\right)
\nonumber \\ &
+\gamma(2\pi)^{3/2}\tilde{C}_R(k_1)
\left \langle 
\hat{p}_{{\bm k}_2}
\hat{v}_{{\bm k}_3}
\right\rangle
\delta \left({\bm k}_1+{\bm k}_4\right)
\nonumber \\ &
+\gamma(2\pi)^{3/2}\tilde{C}_R(k_1)
\left \langle 
\hat{v}_{{\bm k}_3}
\hat{p}_{{\bm k}_4}
\right\rangle
\delta \left({\bm k}_1+{\bm k}_2\right),
\\
\frac{\dd  }{\dd  \eta} \left\langle \hat{p}_{{\bm k}_1}
\hat{p}_{{\bm k}_2}\hat{p}_{{\bm k}_3}\hat{v}_{{\bm k}_4}\right\rangle &=
\left \langle 
\hat{p}_{{\bm k}_1}
\hat{p}_{{\bm k}_2}
\hat{p}_{{\bm k}_3}
\hat{p}_{{\bm k}_4}
\right\rangle
-\omega^2(k_3)
\left \langle 
\hat{p}_{{\bm k}_1}
\hat{p}_{{\bm k}_2}
\hat{v}_{{\bm k}_3}
\hat{v}_{{\bm k}_4}
\right\rangle
-\omega^2(k_2)
\left \langle 
\hat{p}_{{\bm k}_1}
\hat{v}_{{\bm k}_2}
\hat{p}_{{\bm k}_3}
\hat{v}_{{\bm k}_4}
\right\rangle
\nonumber \\
& -\omega^2(k_1)
\left \langle 
\hat{v}_{{\bm k}_1}
\hat{p}_{{\bm k}_2}
\hat{p}_{{\bm k}_3}
\hat{v}_{{\bm k}_4}
\right\rangle
+\gamma(2\pi)^{3/2}\tilde{C}_R(k_1)
\left \langle 
\hat{p}_{{\bm k}_1}
\hat{v}_{{\bm k}_4}
\right\rangle
\delta \left({\bm k}_2+{\bm k}_3\right)
\nonumber \\ &
+\gamma(2\pi)^{3/2}\tilde{C}_R(k_1)
\left \langle 
\hat{p}_{{\bm k}_2}
\hat{v}_{{\bm k}_4}
\right\rangle
\delta \left({\bm k}_1+{\bm k}_3\right)
\nonumber \\ &
+\gamma(2\pi)^{3/2}\tilde{C}_R(k_1)
\left \langle 
\hat{p}_{{\bm k}_3}
\hat{v}_{{\bm k}_4}
\right\rangle
\delta \left({\bm k}_1+{\bm k}_2\right),
\\
\frac{\dd  }{\dd  \eta} \left\langle \hat{p}_{{\bm k}_1}
\hat{p}_{{\bm k}_2}\hat{p}_{{\bm k}_3}\hat{p}_{{\bm k}_4}\right\rangle &=
-\omega^2(k_4)
\left \langle 
\hat{p}_{{\bm k}_1}
\hat{p}_{{\bm k}_2}
\hat{p}_{{\bm k}_3}
\hat{v}_{{\bm k}_4}
\right\rangle
-\omega^2(k_3)
\left \langle 
\hat{p}_{{\bm k}_1}
\hat{p}_{{\bm k}_2}
\hat{v}_{{\bm k}_3}
\hat{p}_{{\bm k}_4}
\right\rangle
\nonumber \\ 
& -\omega^2(k_2)
\left \langle 
\hat{p}_{{\bm k}_1}
\hat{v}_{{\bm k}_2}
\hat{p}_{{\bm k}_3}
\hat{p}_{{\bm k}_4}
\right\rangle
-\omega^2(k_1)
\left \langle 
\hat{v}_{{\bm k}_1}
\hat{p}_{{\bm k}_2}
\hat{p}_{{\bm k}_3}
\hat{p}_{{\bm k}_4}
\right\rangle 
\nonumber \\ &
+\gamma(2\pi)^{3/2}\tilde{C}_R(k_3)
\left \langle 
\hat{p}_{{\bm k}_1}
\hat{p}_{{\bm k}_2}
\right\rangle
\delta \left({\bm k}_3+{\bm k}_4\right)
\nonumber \\ &
+\gamma(2\pi)^{3/2}\tilde{C}_R(k_2)
\left \langle 
\hat{p}_{{\bm k}_1}
\hat{p}_{{\bm k}_3}
\right\rangle
\delta \left({\bm k}_2+{\bm k}_4\right)
\nonumber \\ &
+\gamma(2\pi)^{3/2}\tilde{C}_R(k_1)
\left \langle 
\hat{p}_{{\bm k}_2}
\hat{p}_{{\bm k}_3}
\right\rangle
\delta \left({\bm k}_1+{\bm k}_4\right)
\nonumber \\ &
+\gamma(2\pi)^{3/2}\tilde{C}_R(k_2)
\left \langle 
\hat{p}_{{\bm k}_1}
\hat{p}_{{\bm k}_4}
\right\rangle
\delta \left({\bm k}_2+{\bm k}_4\right)
\nonumber \\ &
+\gamma(2\pi)^{3/2}\tilde{C}_R(k_1)
\left \langle 
\hat{p}_{{\bm k}_2}
\hat{p}_{{\bm k}_4}
\right\rangle
\delta \left({\bm k}_1+{\bm k}_3\right)
\nonumber \\ &
+\gamma(2\pi)^{3/2}\tilde{C}_R(k_1)
\left \langle 
\hat{p}_{{\bm k}_3}
\hat{p}_{{\bm k}_4}
\right\rangle
\delta \left({\bm k}_1+{\bm k}_2\right).
\end{align}
We notice that, contrary to the three-point correlators, the above
equations contain terms proportional to $\gamma $. This is because these correlators contain disconnected diagrams, \ie products of power spectra which we have shown are affected by the environment, see \Eq{eq:pplinear}. Non Gaussianities are however not present in the \emph{connected} four-point correlators that one obtains from subtracting the contribution of Wick theorem (for instance, 
$\left \langle \hat{v}_{{\bm k}_1}\hat{p}_{{\bm k}_2}
\hat{p}_{{\bm k}_3}\hat{v}_{{\bm k}_4}\right\rangle_\uc =
 \left \langle \hat{v}_{{\bm k}_1}\hat{p}_{{\bm k}_2}\hat{p}_{{\bm k}_3}\hat{v}_{{\bm k}_4}\right\rangle
  - \left \langle \hat{v}_{{\bm k}_1}\hat{p}_{{\bm k}_2} \right\rangle \left \langle\hat{p}_{{\bm k}_3}\hat{v}_{{\bm k}_4}\right\rangle
 - \left \langle   \hat{v}_{{\bm k}_1}  \hat{p}_{{\bm k}_3} \right\rangle  \left \langle \hat{p}_{{\bm k}_2}\hat{v}_{{\bm k}_4}   \right\rangle
  - \left \langle \hat{v}_{{\bm k}_1} \hat{v}_{{\bm k}_4}   \right\rangle  \left \langle \hat{p}_{{\bm k}_2}\hat{p}_{{\bm k}_3}  \right\rangle
   $, where the index $\mathrm{c}$ denotes the connected part of the correlator). The evolution equations for the connected four-point correlators can be derived from the previous set of equations making use of 
\Eqs{eq:vvlinear}-(\ref{eq:pplinear}), and one can check that the corresponding system
of equations no longer contains terms proportional to $\gamma$ (one may have
noticed that the terms proportional to $\gamma $ have the same form in
the equations for the two-point and for the four-point correlators). This is again consistent with the fact
that the density matrix of the system remains Gaussian.

\section{Correlators for quadratic interactions}
\label{sec:appquadratic}
In this section, we derive the equations for the three- and four-point
correlation functions in the case where the interaction between the
system and the environment is proportional to the square of the
Mukhanov-Sasaki variable. Let us recall that the equations for the two-point correlators have been established in
\Ref{Martin:2018zbe} and are given by
\begin{align}
\label{eq:vvquadratic}
\frac{\dd  }{\dd  \eta} \left\langle \hat{v}_{{\bm k}_1}
\hat{v}_{{\bm k}_2}\right\rangle &=\left \langle 
\hat{v}_{{\bm k}_1}
\hat{p}_{{\bm k}_2}
+\hat{p}_{{\bm k}_1}
\hat{v}_{{\bm k}_2}
\right\rangle, \\
\label{eq:vpquadratic}
\frac{\dd  }{\dd  \eta} \left\langle \hat{v}_{{\bm k}_1}
\hat{p}_{{\bm k}_2}\right\rangle &=\left \langle 
\hat{p}_{{\bm k}_1}
\hat{p}_{{\bm k}_2}\right \rangle
-\omega^2(k_2)\left\langle
\hat{v}_{{\bm k}_1}
\hat{v}_{{\bm k}_2}
\right\rangle,\\
\label{eq:pvquadratic}
\frac{\dd  }{\dd  \eta} \left\langle \hat{p}_{{\bm k}_1}
\hat{v}_{{\bm k}_2}\right\rangle &=\left \langle 
\hat{p}_{{\bm k}_1}
\hat{p}_{{\bm k}_2}\right \rangle
-\omega^2(k_1)\left\langle
\hat{v}_{{\bm k}_1}
\hat{v}_{{\bm k}_2}
\right\rangle,\\
\label{eq:ppquadratic}
\frac{\dd  }{\dd  \eta} \left\langle \hat{p}_{{\bm k}_1}
\hat{p}_{{\bm k}_2}\right\rangle &=
-\omega^2(k_2)\left\langle\hat{p}_{{\bm k}_1}
\hat{v}_{{\bm k}_2}\right \rangle
-\omega^2(k_1)\left\langle
\hat{v}_{{\bm k}_1}
\hat{p}_{{\bm k}_2}
\right\rangle
\\ &
+\frac{4\gamma}{(2\pi)^{3/2}}
\int \dd ^3 {\bm k}\, \tilde{C}_{R}(\bm k)
\left \langle \hat{v}_{{\bm k}+{\bm k}_1}
\hat{v}_{-{\bm k}+{\bm k}_2}\right\rangle.
\end{align}

\subsection{Three-point correlators}
\label{subsec:threequadratic}

Using the Lindblad equation~(\ref{eq:Lindblad:mean:Fourrier}), the
equations controlling the evolution of the three-point correlators can
be expressed as
\begin{align}
\label{eq:3points_quadratic}
\frac{\dd  }{\dd  \eta} \left\langle \hat{v}_{{\bm k}_1}
\hat{v}_{{\bm k}_2}\hat{v}_{{\bm k}_3}\right\rangle &=
\left \langle 
\hat{v}_{{\bm k}_1}
\hat{v}_{{\bm k}_2}
\hat{p}_{{\bm k}_3}\right\rangle
+\left \langle 
\hat{v}_{{\bm k}_1}
\hat{p}_{{\bm k}_2}
\hat{v}_{{\bm k}_3}\right\rangle
+
\left \langle 
\hat{p}_{{\bm k}_1}
\hat{v}_{{\bm k}_2}
\hat{v}_{{\bm k}_3}\right\rangle, \\
\frac{\dd  }{\dd  \eta} \left\langle \hat{v}_{{\bm k}_1}
\hat{v}_{{\bm k}_2}\hat{p}_{{\bm k}_3}\right\rangle &=
\left \langle 
\hat{v}_{{\bm k}_1}
\hat{p}_{{\bm k}_2}
\hat{p}_{{\bm k}_3}\right\rangle
+\left \langle 
\hat{p}_{{\bm k}_1}
\hat{v}_{{\bm k}_2}
\hat{p}_{{\bm k}_3}\right\rangle
-\omega^2(k_3)
\left \langle 
\hat{v}_{{\bm k}_1}
\hat{v}_{{\bm k}_2}
\hat{v}_{{\bm k}_3}\right\rangle, \\
\frac{\dd  }{\dd  \eta} \left\langle \hat{v}_{{\bm k}_1}
\hat{p}_{{\bm k}_2}\hat{v}_{{\bm k}_3}\right\rangle &=
\left \langle 
\hat{v}_{{\bm k}_1}
\hat{p}_{{\bm k}_2}
\hat{p}_{{\bm k}_3}\right\rangle
+\left \langle 
\hat{p}_{{\bm k}_1}
\hat{p}_{{\bm k}_2}
\hat{v}_{{\bm k}_3}\right\rangle
-\omega^2(k_2)
\left \langle 
\hat{v}_{{\bm k}_1}
\hat{v}_{{\bm k}_2}
\hat{v}_{{\bm k}_3}\right\rangle, \\
\frac{\dd  }{\dd  \eta} \left\langle \hat{p}_{{\bm k}_1}
\hat{v}_{{\bm k}_2}\hat{v}_{{\bm k}_3}\right\rangle &=
\left \langle 
\hat{p}_{{\bm k}_1}
\hat{v}_{{\bm k}_2}
\hat{p}_{{\bm k}_3}\right\rangle
+\left \langle 
\hat{p}_{{\bm k}_1}
\hat{p}_{{\bm k}_2}
\hat{v}_{{\bm k}_3}\right\rangle
-\omega^2(k_1)
\left \langle 
\hat{v}_{{\bm k}_1}
\hat{v}_{{\bm k}_2}
\hat{v}_{{\bm k}_3}\right\rangle, \\
\frac{\dd  }{\dd  \eta} \left\langle \hat{v}_{{\bm k}_1}
\hat{p}_{{\bm k}_2}\hat{p}_{{\bm k}_3}\right\rangle &=
\left \langle 
\hat{p}_{{\bm k}_1}
\hat{p}_{{\bm k}_2}
\hat{p}_{{\bm k}_3}\right\rangle
-\omega^2(k_3)
\left \langle 
\hat{v}_{{\bm k}_1}
\hat{p}_{{\bm k}_2}
\hat{v}_{{\bm k}_3}\right\rangle
-\omega^2(k_2)
\left \langle 
\hat{v}_{{\bm k}_1}
\hat{v}_{{\bm k}_2}
\hat{p}_{{\bm k}_3}\right\rangle
\nonumber \\ &
+\frac{4\gamma}{(2\pi)^{3/2}}\int {\mathrm d}^3 {\bm k}
\, \tilde{C}_R(k)
\left \langle 
\hat{v}_{{\bm k}_1}
\hat{v}_{{\bm k}_2-{\bm k}}
\hat{v}_{{\bm k}+{\bm k}_3}\right\rangle, 
\\
\frac{\dd  }{\dd  \eta} \left\langle \hat{p}_{{\bm k}_1}
\hat{v}_{{\bm k}_2}\hat{p}_{{\bm k}_3}\right\rangle &=
\left \langle 
\hat{p}_{{\bm k}_1}
\hat{p}_{{\bm k}_2}
\hat{p}_{{\bm k}_3}\right\rangle
-\omega^2(k_1)
\left \langle 
\hat{v}_{{\bm k}_1}
\hat{v}_{{\bm k}_2}
\hat{p}_{{\bm k}_3}\right\rangle
-\omega^2(k_3)
\left \langle 
\hat{p}_{{\bm k}_1}
\hat{v}_{{\bm k}_2}
\hat{v}_{{\bm k}_3}\right\rangle
\nonumber \\ &
+\frac{4\gamma}{(2\pi)^{3/2}}\int {\mathrm d}^3 {\bm k}
\, \tilde{C}_R(k)
\left \langle 
\hat{v}_{{\bm k}_1-{\bm k}}
\hat{v}_{{\bm k}_2}
\hat{v}_{{\bm k}+{\bm k}_3}\right\rangle, 
\\
\frac{\dd  }{\dd  \eta} \left\langle \hat{p}_{{\bm k}_1}
\hat{p}_{{\bm k}_2}\hat{v}_{{\bm k}_3}\right\rangle &=
\left \langle 
\hat{p}_{{\bm k}_1}
\hat{p}_{{\bm k}_2}
\hat{p}_{{\bm k}_3}\right\rangle
-\omega^2(k_2)
\left \langle 
\hat{p}_{{\bm k}_1}
\hat{v}_{{\bm k}_2}
\hat{v}_{{\bm k}_3}\right\rangle
-\omega^2(k_1)
\left \langle 
\hat{v}_{{\bm k}_1}
\hat{p}_{{\bm k}_2}
\hat{v}_{{\bm k}_3}\right\rangle
\nonumber \\ &
+\frac{4\gamma}{(2\pi)^{3/2}}\int {\mathrm d}^3 {\bm k}
\, \tilde{C}_R(k)
\left \langle 
\hat{v}_{{\bm k}_1-{\bm k}}
\hat{v}_{{\bm k}+{\bm k}_2}
\hat{v}_{{\bm k}_3}\right\rangle, 
\\
\frac{\dd  }{\dd  \eta} \left\langle \hat{p}_{{\bm k}_1}
\hat{p}_{{\bm k}_2}\hat{v}_{{\bm k}_3}\right\rangle &=
-\omega^2(k_1)
\left \langle 
\hat{v}_{{\bm k}_1}
\hat{p}_{{\bm k}_2}
\hat{p}_{{\bm k}_3}\right\rangle
-\omega^2(k_2)
\left \langle 
\hat{p}_{{\bm k}_1}
\hat{v}_{{\bm k}_2}
\hat{p}_{{\bm k}_3}\right\rangle
-\omega^2(k_3)
\left \langle 
\hat{p}_{{\bm k}_1}
\hat{p}_{{\bm k}_2}
\hat{v}_{{\bm k}_3}\right\rangle
\nonumber \\ &
+\frac{4\gamma}{(2\pi)^{3/2}}\int {\mathrm d}^3 {\bm k}
\, \tilde{C}_R(k)\bigl(
\left \langle 
\hat{p}_{{\bm k}_1}
\hat{v}_{{\bm k}_2-{\bm k}}
\hat{v}_{{\bm k}+{\bm k}_3}\right\rangle
+
\left \langle 
\hat{v}_{{\bm k}_1-{\bm k}}
\hat{p}_{{\bm k}_2}
\hat{v}_{{\bm k}+{\bm k}_3}\right\rangle
\nonumber \\ &
+
\left \langle 
\hat{v}_{{\bm k}_1+{\bm k}}
\hat{v}_{{\bm k}_2-{\bm k}}
\hat{p}_{{\bm k}_3}\right\rangle
\bigr)
.
\end{align}
We notice that the above system of equations contain terms
proportional to $\gamma $. As is typical for quadratic interactions, see \Eqs{eq:vvquadratic}-(\ref{eq:ppquadratic}),
and contrary to the case of linear interactions, see \Eqs{eq:vvlinear}-(\ref{eq:pplinear}), they involve integrals over momentum. However, those terms are also all expressed in terms of
three-point correlators [in the very same way, in the case of the
equations~(\ref{eq:vvquadratic})-(\ref{eq:ppquadratic}) for the two-point correlation functions, the term
proportional to $\gamma $ is proportional to the power
spectrum]. Since these corrections must be evaluated in the free
theory (we recall that the Lindblad equation is established
perturbatively in the interaction strength and is valid at first order in $\gamma $ only), this implies that they all vanish at
leading order in $\gamma$. As already mentioned in the main text, we
conclude that the Lindblad equation does not lead to a non-vanishing bispectrum in that case.

\subsection{Four-point correlators}
\label{subsec:fourquadratic}

This is why one needs to calculate the four-point
correlators (namely the trispectrum) if one wants to exhibit
non Gaussianities. As already discussed in \App{sec:applinear},
the fact that these equations contain terms proportional to $\gamma $
does not necessarily imply that non Gaussianities are present. For
this reason, we now directly proceed to the calculation of the
connected part of the four-point correlators. Straightforward but lengthy calculations lead to the
following expressions
\begin{align}
\label{eq:vvvv}
\frac{\dd  }{\dd  \eta} \left\langle \hat{v}_{{\bm k}_1}
\hat{v}_{{\bm k}_2}\hat{v}_{{\bm k}_3}\hat{v}_{{\bm k}_4}
\right\rangle_\uc &=
\left \langle 
\hat{v}_{{\bm k}_1}
\hat{v}_{{\bm k}_2}
\hat{v}_{{\bm k}_3}
\hat{p}_{{\bm k}_4}
\right\rangle_\uc
+
\left \langle 
\hat{v}_{{\bm k}_1}
\hat{v}_{{\bm k}_2}
\hat{p}_{{\bm k}_3}
\hat{v}_{{\bm k}_4}
\right\rangle_\uc
+
\left \langle 
\hat{v}_{{\bm k}_1}
\hat{p}_{{\bm k}_2}
\hat{v}_{{\bm k}_3}
\hat{v}_{{\bm k}_4}
\right\rangle_\uc
\nonumber \\ &
+
\left \langle 
\hat{p}_{{\bm k}_1}
\hat{v}_{{\bm k}_2}
\hat{v}_{{\bm k}_3}
\hat{v}_{{\bm k}_4}
\right\rangle_\uc, 
\\
\label{eq:vvvp1}
\frac{\dd  }{\dd  \eta} \left\langle \hat{v}_{{\bm k}_1}
\hat{v}_{{\bm k}_2}\hat{v}_{{\bm k}_3}\hat{p}_{{\bm k}_4}
\right\rangle_\uc &=
\left \langle 
\hat{v}_{{\bm k}_1}
\hat{v}_{{\bm k}_2}
\hat{p}_{{\bm k}_3}
\hat{p}_{{\bm k}_4}
\right\rangle_\uc
+
\left \langle 
\hat{v}_{{\bm k}_1}
\hat{p}_{{\bm k}_2}
\hat{v}_{{\bm k}_3}
\hat{p}_{{\bm k}_4}
\right\rangle_\uc
+
\left \langle 
\hat{p}_{{\bm k}_1}
\hat{v}_{{\bm k}_2}
\hat{v}_{{\bm k}_3}
\hat{p}_{{\bm k}_4}
\right\rangle_\uc
\nonumber \\
& -\omega^2(k_4)
\left \langle 
\hat{v}_{{\bm k}_1}
\hat{v}_{{\bm k}_2}
\hat{v}_{{\bm k}_3}
\hat{v}_{{\bm k}_4}
\right\rangle_\uc, 
\\
\label{eq:vvvp2}
\frac{\dd  }{\dd  \eta} \left\langle \hat{v}_{{\bm k}_1}
\hat{v}_{{\bm k}_2}\hat{p}_{{\bm k}_3}\hat{v}_{{\bm k}_4}
\right\rangle_\uc &=
\left \langle 
\hat{p}_{{\bm k}_1}
\hat{v}_{{\bm k}_2}
\hat{p}_{{\bm k}_3}
\hat{v}_{{\bm k}_4}
\right\rangle_\uc
+
\left \langle 
\hat{v}_{{\bm k}_1}
\hat{v}_{{\bm k}_2}
\hat{p}_{{\bm k}_3}
\hat{p}_{{\bm k}_4}
\right\rangle_\uc
+
\left \langle 
\hat{v}_{{\bm k}_1}
\hat{p}_{{\bm k}_2}
\hat{p}_{{\bm k}_3}
\hat{v}_{{\bm k}_4}
\right\rangle_\uc
\nonumber \\
& -\omega^2(k_3)
\left \langle 
\hat{v}_{{\bm k}_1}
\hat{v}_{{\bm k}_2}
\hat{v}_{{\bm k}_3}
\hat{v}_{{\bm k}_4}
\right\rangle_\uc, 
\\
\label{eq:vvvp3}
\frac{\dd  }{\dd  \eta} \left\langle \hat{v}_{{\bm k}_1}
\hat{p}_{{\bm k}_2}\hat{v}_{{\bm k}_3}\hat{v}_{{\bm k}_4}
\right\rangle_\uc &=
\left \langle 
\hat{v}_{{\bm k}_1}
\hat{p}_{{\bm k}_2}
\hat{v}_{{\bm k}_3}
\hat{p}_{{\bm k}_4}
\right\rangle_\uc
+
\left \langle 
\hat{v}_{{\bm k}_1}
\hat{p}_{{\bm k}_2}
\hat{p}_{{\bm k}_3}
\hat{v}_{{\bm k}_4}
\right\rangle_\uc
+
\left \langle 
\hat{p}_{{\bm k}_1}
\hat{p}_{{\bm k}_2}
\hat{v}_{{\bm k}_3}
\hat{v}_{{\bm k}_4}
\right\rangle_\uc
\nonumber \\
& -\omega^2(k_2)
\left \langle 
\hat{v}_{{\bm k}_1}
\hat{v}_{{\bm k}_2}
\hat{v}_{{\bm k}_3}
\hat{v}_{{\bm k}_4}
\right\rangle_\uc, 
\\
\label{eq:vvvp4}
\frac{\dd  }{\dd  \eta} \left\langle \hat{p}_{{\bm k}_1}
\hat{v}_{{\bm k}_2}\hat{v}_{{\bm k}_3}\hat{v}_{{\bm k}_4}
\right\rangle_\uc &=
\left \langle 
\hat{p}_{{\bm k}_1}
\hat{v}_{{\bm k}_2}
\hat{v}_{{\bm k}_3}
\hat{p}_{{\bm k}_4}
\right\rangle_\uc
+
\left \langle 
\hat{p}_{{\bm k}_1}
\hat{v}_{{\bm k}_2}
\hat{p}_{{\bm k}_3}
\hat{v}_{{\bm k}_4}
\right\rangle_\uc
+
\left \langle 
\hat{p}_{{\bm k}_1}
\hat{p}_{{\bm k}_2}
\hat{v}_{{\bm k}_3}
\hat{v}_{{\bm k}_4}
\right\rangle_\uc
\nonumber \\
& -\omega^2(k_1)
\left \langle 
\hat{v}_{{\bm k}_1}
\hat{v}_{{\bm k}_2}
\hat{v}_{{\bm k}_3}
\hat{v}_{{\bm k}_4}
\right\rangle_\uc, 
\\
\label{eq:vvpp1}
\frac{\dd  }{\dd  \eta} \left\langle \hat{v}_{{\bm k}_1}
\hat{v}_{{\bm k}_2}\hat{p}_{{\bm k}_3}\hat{p}_{{\bm k}_4}
\right\rangle_\uc &=
\left \langle 
\hat{v}_{{\bm k}_1}
\hat{p}_{{\bm k}_2}
\hat{p}_{{\bm k}_3}
\hat{p}_{{\bm k}_4}
\right\rangle_\uc
+
\left \langle 
\hat{p}_{{\bm k}_1}
\hat{v}_{{\bm k}_2}
\hat{p}_{{\bm k}_3}
\hat{p}_{{\bm k}_4}
\right\rangle_\uc
\nonumber \\ &
-\omega^2(k_4)
\left \langle 
\hat{v}_{{\bm k}_1}
\hat{v}_{{\bm k}_2}
\hat{p}_{{\bm k}_3}
\hat{v}_{{\bm k}_4}
\right\rangle_\uc
 -\omega^2(k_3)
\left \langle 
\hat{v}_{{\bm k}_1}
\hat{v}_{{\bm k}_2}
\hat{v}_{{\bm k}_3}
\hat{p}_{{\bm k}_4}
\right\rangle_\uc
\nonumber \\ &
+\frac{4\gamma}{(2\pi)^{3/2}}\int {\mathrm d}^3 {\bm k}
\, \tilde{C}_R(k)
\left \langle 
\hat{v}_{{\bm k}_1}
\hat{v}_{{\bm k}_2}
\hat{v}_{{\bm k}_3-{\bm k}}
\hat{v}_{{\bm k}_4+{\bm k}}
\right\rangle_\uc
\nonumber \\ &
+\frac{4\gamma}{(2\pi)^{3/2}}
P_{vv}\left({\bm k}_1\right)P_{vv}\left({\bm k}_2\right)
\left[\tilde{C}_R\left(\left \vert {\bm k}_1+{\bm k}_3\right \vert \right)
+\tilde{C}_R\left(\left\vert {\bm k}_1+{\bm k}_4\right \vert \right)\right]
\nonumber \\ & \times
\delta\left({\bm k}_1+{\bm k}_2+{\bm k}_3+{\bm k}_4\right)
,
\\
\label{eq:vvpp2}
\frac{\dd  }{\dd  \eta} \left\langle \hat{v}_{{\bm k}_1}
\hat{p}_{{\bm k}_2}\hat{v}_{{\bm k}_3}\hat{p}_{{\bm k}_4}
\right\rangle_\uc &=
\left \langle 
\hat{v}_{{\bm k}_1}
\hat{p}_{{\bm k}_2}
\hat{p}_{{\bm k}_3}
\hat{p}_{{\bm k}_4}
\right\rangle_\uc
+
\left \langle 
\hat{p}_{{\bm k}_1}
\hat{p}_{{\bm k}_2}
\hat{v}_{{\bm k}_3}
\hat{p}_{{\bm k}_4}
\right\rangle_\uc
\nonumber \\ &
-\omega^2(k_4)
\left \langle 
\hat{v}_{{\bm k}_1}
\hat{p}_{{\bm k}_2}
\hat{v}_{{\bm k}_3}
\hat{v}_{{\bm k}_4}
\right\rangle_\uc
-\omega^2(k_2)
\left \langle 
\hat{v}_{{\bm k}_1}
\hat{v}_{{\bm k}_2}
\hat{v}_{{\bm k}_3}
\hat{p}_{{\bm k}_4}
\right\rangle_\uc
\nonumber \\ &
+\frac{4\gamma}{(2\pi)^{3/2}}\int {\mathrm d}^3 {\bm k}
\, \tilde{C}_R(k)
\left \langle 
\hat{v}_{{\bm k}_1}
\hat{v}_{{\bm k}_2-{\bm k}}
\hat{v}_{{\bm k}_3}
\hat{v}_{{\bm k}_4+{\bm k}}
\right\rangle_\uc
\nonumber \\ &
+\frac{4\gamma}{(2\pi)^{3/2}}
P_{vv}\left(\bm{k}_1\right)P_{vv}\left( \bm{k}_3\right)
\left[\tilde{C}_R\left(\left \vert \bm{k}_1+\bm{k}_2\right \vert \right)
+\tilde{C}_R\left(\left \vert \bm{k}_1+\bm{k}_4\right \vert \right) 
\right]
\nonumber \\ & \times
\delta\left({\bm k}_1+{\bm k}_2+{\bm k}_3+{\bm k}_4\right)
,
\\
\label{eq:vvpp3}
\frac{\dd  }{\dd  \eta} \left\langle \hat{p}_{{\bm k}_1}
\hat{v}_{{\bm k}_2}\hat{v}_{{\bm k}_3}\hat{p}_{{\bm k}_4}
\right\rangle_\uc &=
\left \langle 
\hat{p}_{{\bm k}_1}
\hat{v}_{{\bm k}_2}
\hat{p}_{{\bm k}_3}
\hat{p}_{{\bm k}_4}
\right\rangle_\uc
+
\left \langle 
\hat{p}_{{\bm k}_1}
\hat{p}_{{\bm k}_2}
\hat{v}_{{\bm k}_3}
\hat{p}_{{\bm k}_4}
\right\rangle_\uc
\nonumber \\ &
-\omega^2(k_4)
\left \langle 
\hat{p}_{{\bm k}_1}
\hat{v}_{{\bm k}_2}
\hat{v}_{{\bm k}_3}
\hat{v}_{{\bm k}_4}
\right\rangle_\uc
-\omega^2(k_1)
\left \langle 
\hat{v}_{{\bm k}_1}
\hat{v}_{{\bm k}_2}
\hat{v}_{{\bm k}_3}
\hat{p}_{{\bm k}_4}
\right\rangle_\uc
\nonumber \\ &
+\frac{4\gamma}{(2\pi)^{3/2}}\int {\mathrm d}^3 {\bm k}
\, \tilde{C}_R(k)
\left \langle 
\hat{v}_{{\bm k}_1-{\bm k}}
\hat{v}_{{\bm k}_2}
\hat{v}_{{\bm k}_3}
\hat{v}_{{\bm k}_4+{\bm k}}
\right\rangle_\uc
\nonumber \\ & 
+\frac{4\gamma}{(2\pi)^{3/2}}
P_{vv}\left(\bm{k}_2\right)P_{vv}\left( \bm{k}_3\right)
\left[\tilde{C}_R\left(\left \vert \bm{k}_1+\bm{k}_2\right\vert \right)
+\tilde{C}_R\left(\left\vert \bm{k}_1+\bm{k}_3\right\vert \right) 
\right]
\nonumber \\ & \times
\delta\left({\bm k}_1+{\bm k}_2+{\bm k}_3+{\bm k}_4\right)
,
\\
\label{eq:vvpp4}
\frac{\dd  }{\dd  \eta} \left\langle \hat{v}_{{\bm k}_1}
\hat{p}_{{\bm k}_2}\hat{p}_{{\bm k}_3}\hat{v}_{{\bm k}_4}
\right\rangle_\uc &=
\left \langle 
\hat{v}_{{\bm k}_1}
\hat{p}_{{\bm k}_2}
\hat{p}_{{\bm k}_3}
\hat{p}_{{\bm k}_4}
\right\rangle_\uc
+
\left \langle 
\hat{p}_{{\bm k}_1}
\hat{p}_{{\bm k}_2}
\hat{p}_{{\bm k}_3}
\hat{v}_{{\bm k}_4}
\right\rangle_\uc
\nonumber \\ &
-\omega^2(k_3)
\left \langle 
\hat{v}_{{\bm k}_1}
\hat{p}_{{\bm k}_2}
\hat{v}_{{\bm k}_3}
\hat{v}_{{\bm k}_4}
\right\rangle_\uc
-\omega^2(k_2)
\left \langle 
\hat{v}_{{\bm k}_1}
\hat{v}_{{\bm k}_2}
\hat{p}_{{\bm k}_3}
\hat{v}_{{\bm k}_4}
\right\rangle_\uc
\nonumber \\ &
+\frac{4\gamma}{(2\pi)^{3/2}}\int {\mathrm d}^3 {\bm k}
\, \tilde{C}_R(k)
\left \langle 
\hat{v}_{{\bm k}_1}
\hat{v}_{{\bm k}_2-{\bm k}}
\hat{v}_{{\bm k}_3+{\bm k}}
\hat{v}_{{\bm k}_4}
\right\rangle_\uc
\nonumber \\ & 
+\frac{4\gamma}{(2\pi)^{3/2}}
P_{vv}\left(\bm{k}_1\right)P_{vv}\left( \bm{k}_4\right)
\left[\tilde{C}_R\left(\left \vert \bm{k}_1+\bm{k}_2\right \vert \right)
+\tilde{C}_R\left(\left \vert \bm{k}_1+\bm{k}_3\right \vert \right) 
\right]
\nonumber \\ & \times 
\delta\left({\bm k}_1+{\bm k}_2+{\bm k}_3+{\bm k}_4\right)
,
\\
\label{eq:vvpp5}
\frac{\dd  }{\dd  \eta} \left\langle \hat{p}_{{\bm k}_1}
\hat{v}_{{\bm k}_2}\hat{p}_{{\bm k}_3}\hat{v}_{{\bm k}_4}\right\rangle_\uc &=
\left \langle 
\hat{p}_{{\bm k}_1}
\hat{v}_{{\bm k}_2}
\hat{p}_{{\bm k}_3}
\hat{p}_{{\bm k}_4}
\right\rangle_\uc
+
\left \langle 
\hat{p}_{{\bm k}_1}
\hat{p}_{{\bm k}_2}
\hat{p}_{{\bm k}_3}
\hat{v}_{{\bm k}_4}
\right\rangle_\uc
\nonumber \\ &
-\omega^2(k_3)
\left \langle 
\hat{p}_{{\bm k}_1}
\hat{v}_{{\bm k}_2}
\hat{v}_{{\bm k}_3}
\hat{v}_{{\bm k}_4}
\right\rangle_\uc
-\omega^2(k_1)
\left \langle 
\hat{v}_{{\bm k}_1}
\hat{v}_{{\bm k}_2}
\hat{p}_{{\bm k}_3}
\hat{v}_{{\bm k}_4}
\right\rangle_\uc
\nonumber \\ &
+\frac{4\gamma}{(2\pi)^{3/2}}\int {\mathrm d}^3 {\bm k}
\, \tilde{C}_R(k)
\left \langle 
\hat{v}_{{\bm k}_1-{\bm k}}
\hat{v}_{{\bm k}_2}
\hat{v}_{{\bm k}_3+{\bm k}}
\hat{v}_{{\bm k}_4}
\right\rangle_\uc 
\nonumber \\ &
+\frac{4\gamma}{(2\pi)^{3/2}}
P_{vv}\left(\bm{k}_2\right)P_{vv}\left( \bm{k}_4\right)
\left[\tilde{C}_R\left(\left \vert \bm{k}_1+\bm{k}_2\right \vert\right)
+\tilde{C}_R\left(\left \vert \bm{k}_1+\bm{k}_4\right \vert \right) 
\right]
\nonumber \\ & \times
\delta\left({\bm k}_1+{\bm k}_2+{\bm k}_3+{\bm k}_4\right)
,
\\
\label{eq:vvpp6}
\frac{\dd  }{\dd  \eta} \left\langle \hat{p}_{{\bm k}_1}
\hat{p}_{{\bm k}_2}\hat{v}_{{\bm k}_3}\hat{v}_{{\bm k}_4}\right\rangle_\uc &=
\left \langle 
\hat{p}_{{\bm k}_1}
\hat{p}_{{\bm k}_2}
\hat{v}_{{\bm k}_3}
\hat{p}_{{\bm k}_4}
\right\rangle_\uc
+
\left \langle 
\hat{p}_{{\bm k}_1}
\hat{p}_{{\bm k}_2}
\hat{p}_{{\bm k}_3}
\hat{v}_{{\bm k}_4}
\right\rangle_\uc
\nonumber \\ &
-\omega^2(k_2)
\left \langle 
\hat{p}_{{\bm k}_1}
\hat{v}_{{\bm k}_2}
\hat{v}_{{\bm k}_3}
\hat{v}_{{\bm k}_4}
\right\rangle_\uc
-\omega^2(k_1)
\left \langle 
\hat{v}_{{\bm k}_1}
\hat{p}_{{\bm k}_2}
\hat{v}_{{\bm k}_3}
\hat{v}_{{\bm k}_4}
\right\rangle_\uc
\nonumber \\ &
+\frac{4\gamma}{(2\pi)^{3/2}}\int {\mathrm d}^3 {\bm k}
\, \tilde{C}_R(k)
\left \langle 
\hat{v}_{{\bm k}_1-{\bm k}}
\hat{v}_{{\bm k}_2+{\bm k}}
\hat{v}_{{\bm k}_3}
\hat{v}_{{\bm k}_4}
\right\rangle_\uc
\nonumber \\ & 
+\frac{4\gamma}{(2\pi)^{3/2}}
P_{vv}\left(\bm{k}_3\right)P_{vv}\left( \bm{k}_4\right)
\left[\tilde{C}_R\left(\left \vert \bm{k}_1+\bm{k}_3\right \vert \right)
+\tilde{C}_R\left(\left \vert \bm{k}_1+\bm{k}_4\right \vert \right) 
\right]
\nonumber \\ & \times
\delta\left({\bm k}_1+{\bm k}_2+{\bm k}_3+{\bm k}_4\right)
,
\\
\label{eq:vppp1}
\frac{\dd  }{\dd  \eta} \left\langle \hat{v}_{{\bm k}_1}
\hat{p}_{{\bm k}_2}\hat{p}_{{\bm k}_3}\hat{p}_{{\bm k}_4}
\right\rangle_\uc &=
\left \langle 
\hat{p}_{{\bm k}_1}
\hat{p}_{{\bm k}_2}
\hat{p}_{{\bm k}_3}
\hat{p}_{{\bm k}_4}
\right\rangle_\uc
-\omega^2(k_4)
\left \langle 
\hat{v}_{{\bm k}_1}
\hat{p}_{{\bm k}_2}
\hat{p}_{{\bm k}_3}
\hat{v}_{{\bm k}_4}
\right\rangle_\uc
\nonumber \\ &
-\omega^2(k_3)
\left \langle 
\hat{v}_{{\bm k}_1}
\hat{p}_{{\bm k}_2}
\hat{v}_{{\bm k}_3}
\hat{p}_{{\bm k}_4}
\right\rangle_\uc
-\omega^2(k_2)
\left \langle 
\hat{v}_{{\bm k}_1}
\hat{v}_{{\bm k}_2}
\hat{p}_{{\bm k}_3}
\hat{p}_{{\bm k}_4}
\right\rangle_\uc
\nonumber \\ &
+\frac{4\gamma}{(2\pi)^{3/2}}\int {\mathrm d}^3 {\bm k}
\, \tilde{C}_R(k)\Bigl(
\left \langle 
\hat{v}_{{\bm k}_1}
\hat{p}_{{\bm k}_2}
\hat{v}_{{\bm k}_3-{\bm k}}
\hat{v}_{{\bm k}_4+{\bm k}}
\right\rangle_\uc
\nonumber \\ &
+\left \langle 
\hat{v}_{{\bm k}_1}
\hat{v}_{{\bm k}_2-{\bm k}}
\hat{p}_{{\bm k}_3}
\hat{v}_{{\bm k}_4+{\bm k}}
\right\rangle_\uc
+\left \langle 
\hat{v}_{{\bm k}_1}
\hat{v}_{{\bm k}_2-{\bm k}}
\hat{v}_{{\bm k}_3+{\bm k}}
\hat{p}_{{\bm k}_4}
\right\rangle_\uc
\Bigr)
\nonumber \\ & 
+\frac{4\gamma}{\left(2\pi\right)^{3/2}}
\Bigl\lbrace 
P_{vv}\left(\bm{k}_1\right) P_{pv}\left(\bm{k}_2\right)
\left[\tilde{C}_R\left(\left\vert \bm{k}_1+\bm{k}_3\right \vert\right)+
\tilde{C}_R\left(\left \vert \bm{k}_1+\bm{k}_4\right \vert \right)\right]
\nonumber  \\ & 
+
P_{vv}\left(\bm{k}_1\right)
\left[\tilde{C}_R\left(\left \vert \bm{k}_1+\bm{k}_2\right \vert\right) 
P_{pv}\left(\bm{k}_3\right)
+\tilde{C}_R\left(\left \vert \bm{k}_1+\bm{k}_4\right \vert\right) 
P_{vp}\left(\bm{k}_3\right)\right]
\nonumber  \\ & 
+
P_{vv}\left(\bm{k}_1\right) P_{vp}\left(\bm{k}_4\right)
\left[\tilde{C}_R\left(\left \vert \bm{k}_1+\bm{k}_2\right \vert\right)
+\tilde{C}_R\left(\left \vert \bm{k}_1+\bm{k}_3\right \vert\right)\right]
\Bigr\rbrace 
\nonumber \\ & \times
\delta\left(\bm{k}_1+\bm{k}_2+\bm{k}_3+\bm{k}_4\right)
,
\\
\label{eq:vppp2}
\frac{\dd  }{\dd  \eta} \left\langle \hat{p}_{{\bm k}_1}
\hat{v}_{{\bm k}_2}\hat{p}_{{\bm k}_3}\hat{p}_{{\bm k}_4}
\right\rangle_\uc &=
\left \langle 
\hat{p}_{{\bm k}_1}
\hat{p}_{{\bm k}_2}
\hat{p}_{{\bm k}_3}
\hat{p}_{{\bm k}_4}
\right\rangle_\uc
-\omega^2(k_4)
\left \langle 
\hat{p}_{{\bm k}_1}
\hat{v}_{{\bm k}_2}
\hat{p}_{{\bm k}_3}
\hat{v}_{{\bm k}_4}
\right\rangle_\uc
\nonumber \\ &
-\omega^2(k_3)
\left \langle 
\hat{p}_{{\bm k}_1}
\hat{v}_{{\bm k}_2}
\hat{v}_{{\bm k}_3}
\hat{p}_{{\bm k}_4}
\right\rangle_\uc
-\omega^2(k_1)
\left \langle 
\hat{v}_{{\bm k}_1}
\hat{v}_{{\bm k}_2}
\hat{p}_{{\bm k}_3}
\hat{p}_{{\bm k}_4}
\right\rangle_\uc
\nonumber \\ &
+\frac{4\gamma}{(2\pi)^{3/2}}\int {\mathrm d}^3 {\bm k}
\, \tilde{C}_R(k)\Bigl(
\left \langle 
\hat{p}_{{\bm k}_1}
\hat{v}_{{\bm k}_2}
\hat{v}_{{\bm k}_3-{\bm k}}
\hat{v}_{{\bm k}_4+{\bm k}}
\right\rangle_\uc
\nonumber \\ &
+\left \langle 
\hat{v}_{{\bm k}_1-{\bm k}}
\hat{v}_{{\bm k}_2}
\hat{p}_{{\bm k}_3}
\hat{v}_{{\bm k}_4+{\bm k}}
\right\rangle_\uc
+\left \langle 
\hat{v}_{{\bm k}_1-{\bm k}}
\hat{v}_{{\bm k}_2}
\hat{v}_{{\bm k}_3+{\bm k}}
\hat{p}_{{\bm k}_4}
\right\rangle_\uc
\Bigr)
\nonumber \\ &
+\frac{4\gamma}{\left(2\pi\right)^{3/2}}
\Bigl\lbrace
P_{pv}\left(\bm{k}_1\right) P_{vv}\left(\bm{k}_2\right)
\left[\tilde{C}_R\left(\left \vert \bm{k}_1+\bm{k}_3\right \vert \right)+
\tilde{C}_R\left(\left \vert \bm{k}_1+\bm{k}_4\right \vert \right)\right]
\nonumber  \\ &  
+ P_{vv}\left(\bm{k}_2\right)
\left[\tilde{C}_R\left(\left \vert \bm{k}_1+\bm{k}_2\right \vert \right)
P_{pv}\left(\bm{k}_3\right)+
\tilde{C}_R\left(\left \vert \bm{k}_1+\bm{k}_3\right \vert \right)
P_{vp}\left(\bm{k}_3\right)\right]
\nonumber  \\ &  
+P_{vp}\left(\bm{k}_4\right) P_{vv}\left(\bm{k}_2\right)
\left[\tilde{C}_R\left(\left \vert \bm{k}_1+\bm{k}_2\right \vert \right)+
\tilde{C}_R\left(\left \vert \bm{k}_1+\bm{k}_4\right \vert \right)\right]
\Bigr\rbrace 
\nonumber \\ & \times
\delta\left(\bm{k}_1+\bm{k}_2+\bm{k}_3+\bm{k}_4\right)
,
\\
\label{eq:vppp3}
\frac{\dd  }{\dd  \eta} \left\langle \hat{p}_{{\bm k}_1}
\hat{p}_{{\bm k}_2}\hat{v}_{{\bm k}_3}\hat{p}_{{\bm k}_4}
\right\rangle_\uc &=
\left \langle 
\hat{p}_{{\bm k}_1}
\hat{p}_{{\bm k}_2}
\hat{p}_{{\bm k}_3}
\hat{p}_{{\bm k}_4}
\right\rangle_\uc
-\omega^2(k_4)
\left \langle 
\hat{p}_{{\bm k}_1}
\hat{p}_{{\bm k}_2}
\hat{v}_{{\bm k}_3}
\hat{v}_{{\bm k}_4}
\right\rangle_\uc
\nonumber \\ &
-\omega^2(k_2)
\left \langle 
\hat{p}_{{\bm k}_1}
\hat{v}_{{\bm k}_2}
\hat{v}_{{\bm k}_3}
\hat{p}_{{\bm k}_4}
\right\rangle_\uc
 -\omega^2(k_1)
\left \langle 
\hat{v}_{{\bm k}_1}
\hat{p}_{{\bm k}_2}
\hat{v}_{{\bm k}_3}
\hat{p}_{{\bm k}_4}
\right\rangle_\uc
\nonumber \\ &
+\frac{4\gamma}{(2\pi)^{3/2}}\int {\mathrm d}^3 {\bm k}
\, \tilde{C}_R(k)\Bigl(
\left \langle 
\hat{p}_{{\bm k}_1}
\hat{v}_{{\bm k}_2-{\bm k}}
\hat{v}_{{\bm k}_3}
\hat{v}_{{\bm k}_4+{\bm k}}
\right\rangle_\uc
\nonumber \\ &
+\left \langle 
\hat{v}_{{\bm k}_1-{\bm k}}
\hat{p}_{{\bm k}_2}
\hat{v}_{{\bm k}_3}
\hat{v}_{{\bm k}_4+{\bm k}}
\right\rangle_\uc
+\left \langle 
\hat{v}_{{\bm k}_1-{\bm k}}
\hat{v}_{{\bm k}_2+{\bm k}}
\hat{v}_{{\bm k}_3}
\hat{p}_{{\bm k}_4}
\right\rangle_\uc
\Bigr)
\nonumber \\ &
+\frac{4\gamma}{\left(2\pi\right)^{3/2}}
\Bigl\lbrace
P_{pv}\left(\bm{k}_1\right) P_{vv}\left(\bm{k}_3\right)
\left[\tilde{C}_R\left(\left \vert \bm{k}_1+\bm{k}_2\right \vert \right)+
\tilde{C}_R\left(\left \vert \bm{k}_1+\bm{k}_4\right \vert \right)\right]
\nonumber  \\ & 
+P_{vv}\left(\bm{k}_3\right)
\left[\tilde{C}_R\left(\left \vert \bm{k}_1+\bm{k}_2\right \vert \right)
P_{vp}\left(\bm{k}_2\right) +
\tilde{C}_R\left(\left \vert \bm{k}_1+\bm{k}_3\right \vert \right)
P_{pv}\left(\bm{k}_2\right) \right]
\nonumber  \\ & 
+
P_{vp}\left(\bm{k}_4\right) P_{vv}\left(\bm{k}_3\right)
\left[\tilde{C}_R\left(\left \vert \bm{k}_1+\bm{k}_3\right \vert \right)+
\tilde{C}_R\left(\left \vert \bm{k}_1+\bm{k}_4\right \vert \right)\right]
\Bigr\rbrace 
\nonumber \\ & \times
\delta\left(\bm{k}_1+\bm{k}_2+\bm{k}_3+\bm{k}_4\right)
,
\\
\label{eq:vppp4}
\frac{\dd  }{\dd  \eta} \left\langle \hat{p}_{{\bm k}_1}
\hat{p}_{{\bm k}_2}\hat{p}_{{\bm k}_3}\hat{v}_{{\bm k}_4}
\right\rangle_\uc &=
\left \langle 
\hat{p}_{{\bm k}_1}
\hat{p}_{{\bm k}_2}
\hat{p}_{{\bm k}_3}
\hat{p}_{{\bm k}_4}
\right\rangle_\uc
-\omega^2(k_3)
\left \langle 
\hat{p}_{{\bm k}_1}
\hat{p}_{{\bm k}_2}
\hat{v}_{{\bm k}_3}
\hat{v}_{{\bm k}_4}
\right\rangle_\uc
\nonumber \\ &
-\omega^2(k_2)
\left \langle 
\hat{p}_{{\bm k}_1}
\hat{v}_{{\bm k}_2}
\hat{p}_{{\bm k}_3}
\hat{v}_{{\bm k}_4}
\right\rangle_\uc
-\omega^2(k_1)
\left \langle 
\hat{v}_{{\bm k}_1}
\hat{p}_{{\bm k}_2}
\hat{p}_{{\bm k}_3}
\hat{v}_{{\bm k}_4}
\right\rangle_\uc
\nonumber \\ &
+\frac{4\gamma}{(2\pi)^{3/2}}\int {\mathrm d}^3 {\bm k}
\, \tilde{C}_R(k)\Bigl(
\left \langle 
\hat{p}_{{\bm k}_1}
\hat{v}_{{\bm k}_2-{\bm k}}
\hat{v}_{{\bm k}_3+{\bm k}}
\hat{v}_{{\bm k}_4}
\right\rangle_\uc
\nonumber \\ &
+\left \langle 
\hat{v}_{{\bm k}_1-{\bm k}}
\hat{p}_{{\bm k}_2}
\hat{v}_{{\bm k}_3+{\bm k}}
\hat{v}_{{\bm k}_4}
\right\rangle_\uc
+\left \langle 
\hat{v}_{{\bm k}_1-{\bm k}}
\hat{v}_{{\bm k}_2+{\bm k}}
\hat{p}_{{\bm k}_3}
\hat{v}_{{\bm k}_4}
\right\rangle_\uc
\Bigr)
\nonumber \\ &
+\frac{4\gamma}{\left(2\pi\right)^{3/2}}
\Bigl\lbrace
P_{pv}\left(\bm{k}_1\right) P_{vv}\left(\bm{k}_4\right)
\left[\tilde{C}_R\left(\left \vert \bm{k}_1+\bm{k}_2\right \vert \right)+
\tilde{C}_R\left(\left \vert \bm{k}_1+\bm{k}_3\right \vert \right)\right]
\nonumber  \\ & 
+P_{vv}\left(\bm{k}_4\right)
\left[\tilde{C}_R\left(\left \vert \bm{k}_1+\bm{k}_2\right \vert
\right)P_{vp}\left(\bm{k}_2\right)+
\tilde{C}_R\left(\left \vert \bm{k}_1+\bm{k}_4\right \vert \right)
P_{pv}\left(\bm{k}_2\right)\right]
\nonumber  \\ & 
+P_{vp}\left(\bm{k}_3\right) P_{vv}\left(\bm{k}_4\right)
\left[\tilde{C}_R\left(\left \vert \bm{k}_1+\bm{k}_3\right \vert\right)+
\tilde{C}_R\left(\left \vert \bm{k}_1+\bm{k}_4\right \vert \right)\right]
\Bigr\rbrace 
\nonumber \\ & \times
\delta\left(\bm{k}_1+\bm{k}_2+\bm{k}_3+\bm{k}_4\right)
,
\\
\label{eq:pppp}
\frac{\dd  }{\dd  \eta} \left\langle \hat{p}_{{\bm k}_1}
\hat{p}_{{\bm k}_2}\hat{p}_{{\bm k}_3}\hat{p}_{{\bm k}_4}
\right\rangle_\uc &=
-\omega^2(k_4)
\left \langle 
\hat{p}_{{\bm k}_1}
\hat{p}_{{\bm k}_2}
\hat{p}_{{\bm k}_3}
\hat{v}_{{\bm k}_4}
\right\rangle_\uc
-\omega^2(k_3)
\left \langle 
\hat{p}_{{\bm k}_1}
\hat{p}_{{\bm k}_2}
\hat{v}_{{\bm k}_3}
\hat{p}_{{\bm k}_4}
\right\rangle_\uc
\nonumber \\ 
& -\omega^2(k_2)
\left \langle 
\hat{p}_{{\bm k}_1}
\hat{v}_{{\bm k}_2}
\hat{p}_{{\bm k}_3}
\hat{p}_{{\bm k}_4}
\right\rangle_\uc
-\omega^2(k_1)
\left \langle 
\hat{v}_{{\bm k}_1}
\hat{p}_{{\bm k}_2}
\hat{p}_{{\bm k}_3}
\hat{p}_{{\bm k}_4}
\right\rangle_\uc 
\nonumber \\ &
+\frac{4\gamma}{(2\pi)^{3/2}}\int {\mathrm d}^3 {\bm k}
\, \tilde{C}_R(k)\Bigl(
\left \langle 
\hat{p}_{{\bm k}_1}
\hat{p}_{{\bm k}_2}
\hat{v}_{{\bm k}_3-{\bm k}}
\hat{v}_{{\bm k}_4+{\bm k}}
\right\rangle_\uc
+\left \langle 
\hat{p}_{{\bm k}_1}
\hat{v}_{{\bm k}_2-{\bm k}}
\hat{p}_{{\bm k}_3}
\hat{v}_{{\bm k}_4+{\bm k}}
\right\rangle_\uc
\nonumber \\ &
+\left \langle 
\hat{v}_{{\bm k}_1-{\bm k}}
\hat{p}_{{\bm k}_2}
\hat{p}_{{\bm k}_3}
\hat{v}_{{\bm k}_4+{\bm k}}
\right\rangle_\uc
+\left \langle 
\hat{p}_{{\bm k}_1}
\hat{v}_{{\bm k}_2-{\bm k}}
\hat{v}_{{\bm k}_3+{\bm k}}
\hat{p}_{{\bm k}_4}
\right\rangle_\uc
\nonumber \\ &
+\left \langle 
\hat{v}_{{\bm k}_1-{\bm k}}
\hat{p}_{{\bm k}_2}
\hat{v}_{{\bm k}_3+{\bm k}}
\hat{p}_{{\bm k}_4}
\right\rangle_\uc
+\left \langle 
\hat{v}_{{\bm k}_1-{\bm k}}
\hat{v}_{{\bm k}_2+{\bm k}}
\hat{p}_{{\bm k}_3}
\hat{p}_{{\bm k}_4}
\right\rangle_\uc
\Bigr)
\nonumber \\ &
+\frac{4\gamma}{\left(2\pi\right)^{3/2}}
\biggl\lbrace
P_{pv}\left(\bm{k}_1\right) P_{pv}\left(\bm{k}_2\right)
\left[\tilde{C}_R\left(\left \vert \bm{k}_1+\bm{k}_3\right \vert \right)+
\tilde{C}_R\left(\left \vert \bm{k}_1+\bm{k}_4\right \vert \right)\right]
\nonumber  \\ & 
+P_{pv}\left(\bm{k}_1\right) 
\left[\tilde{C}_R\left(\left \vert \bm{k}_1+\bm{k}_2\right \vert \right)
P_{pv}\left(\bm{k}_3\right)+
\tilde{C}_R\left(\left \vert \bm{k}_1+\bm{k}_4\right \vert \right)
P_{vp}\left(\bm{k}_3\right)\right]
\nonumber  \\ & 
+
\tilde{C}_R\left(\left \vert \bm{k}_1+\bm{k}_2\right \vert \right)
P_{pv}\left(\bm{k}_3\right)P_{vp}\left(\bm{k}_2\right) 
+
\tilde{C}_R\left(\left \vert \bm{k}_1+\bm{k}_3\right \vert \right)
P_{vp}\left(\bm{k}_3\right)P_{pv}\left(\bm{k}_2\right) 
\nonumber  \\ & 
+P_{pv}\left(\bm{k}_1\right)P_{vp}\left(\bm{k}_4\right) 
\left[\tilde{C}_R\left(\left \vert \bm{k}_1+\bm{k}_2\right \vert \right)+
\tilde{C}_R\left(\left \vert \bm{k}_1+\bm{k}_3\right \vert\right) \right]
\nonumber  \\ & 
+P_{vp}\left(\bm{k}_4\right)
\left[\tilde{C}_R\left(\left \vert \bm{k}_1+\bm{k}_2\right \vert \right)
P_{vp}\left(\bm{k}_2\right) +
\tilde{C}_R\left(\left \vert \bm{k}_1+\bm{k}_4\right \vert \right)
P_{pv}\left(\bm{k}_2\right) \right]
\nonumber  \\ & 
+
\tilde{C}_R\left(\left \vert \bm{k}_1+\bm{k}_3\right \vert \right)
P_{vp}\left(\bm{k}_3\right)P_{vp}\left(\bm{k}_4\right)  
+
\tilde{C}_R\left(\left \vert \bm{k}_1+\bm{k}_4\right \vert \right)
P_{vp}\left(\bm{k}_4\right)P_{vp}\left(\bm{k}_3\right) 
\biggr\rbrace
\nonumber \\ & \times
\delta\left(\bm{k}_1+\bm{k}_2+\bm{k}_3+\bm{k}_4\right)
\, .
\end{align}
Let us now discuss these formulas. Among these $16$ equations, $5$
(the first five equations) are ``homogeneous'', in the sense that they
do not contain a source term, contrary to the remaining $11$ ones. We
also see that the source terms are always made of two pieces. One
contains an integral over wavenumbers of the correlation function of
the environment times some connected four point correlators and the
other one is directly proportional to the product of two power spectra
times the environmental correlation function times a Dirac delta
function ensuring that the sum of the four wavenumbers is zero. Of
course, these source terms are proportional to $\gamma $. The source
terms containing the integrals must be ignored since the connected
four-point correlators vanish at leading order in $\gamma $. In other
words, keeping these terms would lead to subdominant contributions,
proportional to $\gamma ^2$, while the Lindblad formalism is consistent only at order $\gamma$.
In the following, we call $\mathfrak{S}_i$, for
$i=1,\cdots,11$, the $11 $ sources terms mentioned above. At leading order in $\gamma$ they are given by
\begin{align}
\label{eq:source1}
\mathfrak{S}_1&=
\frac{4\gamma}{(2\pi)^{3/2}}
P_{vv}\left({\bm k}_1\right)P_{vv}\left({\bm k}_2\right)
\left[\tilde{C}_R\left(\left \vert {\bm k}_1+{\bm k}_3\right \vert \right)
+\tilde{C}_R\left(\left \vert {\bm k}_1+{\bm k}_4\right \vert \right)\right]
\delta\left({\bm k}_1+{\bm k}_2+{\bm k}_3+{\bm k}_4\right)
,
\\
\label{eq:source2}
\mathfrak{S}_2&=\frac{4\gamma}{(2\pi)^{3/2}}
P_{vv}\left(\bm{k}_1\right)P_{vv}\left( \bm{k}_3\right)
\left[\tilde{C}_R\left(\left \vert \bm{k}_1+\bm{k}_2\right \vert \right)
+\tilde{C}_R\left(\left \vert \bm{k}_1+\bm{k}_4\right \vert \right) 
\right]\delta\left({\bm k}_1+{\bm k}_2+{\bm k}_3+{\bm k}_4\right)
,
\\
\label{eq:source3}
\mathfrak{S}_3&=
\frac{4\gamma}{(2\pi)^{3/2}}
P_{vv}\left(\bm{k}_2\right)P_{vv}\left( \bm{k}_3\right)
\left[\tilde{C}_R\left(\left \vert \bm{k}_1+\bm{k}_2\right \vert \right)
+\tilde{C}_R\left(\left \vert \bm{k}_1+\bm{k}_3\right \vert \right) 
\right]\delta\left({\bm k}_1+{\bm k}_2+{\bm k}_3+{\bm k}_4\right)
,
\\
\label{eq:source4}
\mathfrak{S}_4&=\frac{4\gamma}{(2\pi)^{3/2}}
P_{vv}\left(\bm{k}_1\right)P_{vv}\left( \bm{k}_4\right)
\left[\tilde{C}_R\left(\left \vert \bm{k}_1+\bm{k}_2\right \vert \right)
+\tilde{C}_R\left(\left \vert \bm{k}_1+\bm{k}_3\right \vert \right) 
\right]\delta\left({\bm k}_1+{\bm k}_2+{\bm k}_3+{\bm k}_4\right)
,
\\
\label{eq:source5}
\mathfrak{S}_5&=\frac{4\gamma}{(2\pi)^{3/2}}
P_{vv}\left(\bm{k}_2\right)P_{vv}\left( \bm{k}_4\right)
\left[\tilde{C}_R\left(\left \vert \bm{k}_1+\bm{k}_2\right \vert \right)
+\tilde{C}_R\left(\left \vert \bm{k}_1+\bm{k}_4\right\vert \right) 
\right]\delta\left({\bm k}_1+{\bm k}_2+{\bm k}_3+{\bm k}_4\right)
,
\\
\label{eq:source6}
\mathfrak{S}_6&=\frac{4\gamma}{(2\pi)^{3/2}}
P_{vv}\left(\bm{k}_3\right)P_{vv}\left( \bm{k}_4\right)
\left[\tilde{C}_R\left(\left \vert \bm{k}_1+\bm{k}_3\right \vert \right)
+\tilde{C}_R\left(\left \vert \bm{k}_1+\bm{k}_4\right \vert \right) 
\right]\delta\left({\bm k}_1+{\bm k}_2+{\bm k}_3+{\bm k}_4\right)
,
\\
\label{eq:source7}
\mathfrak{S}_7&=\frac{4\gamma}{\left(2\pi\right)^{3/2}}
\left\lbrace
P_{vv}\left(\bm{k}_1\right) P_{pv}\left(\bm{k}_2\right)
\left[\tilde{C}_R\left(\left \vert \bm{k}_1+\bm{k}_3\right \vert\right)+
\tilde{C}_R\left(\left \vert \bm{k}_1+\bm{k}_4\right \vert \right)\right]
\nonumber \right. \\ & \left.
+
P_{vv}\left(\bm{k}_1\right)
\left[\tilde{C}_R\left(\left \vert \bm{k}_1+\bm{k}_2\right \vert \right) 
P_{pv}\left(\bm{k}_3\right)
+\tilde{C}_R\left(\left \vert \bm{k}_1+\bm{k}_4\right \vert \right) 
P_{vp}\left(\bm{k}_3\right)\right]
\nonumber \right. \\ & \left.
+
P_{vv}\left(\bm{k}_1\right) P_{vp}\left(\bm{k}_4\right)
\left[\tilde{C}_R\left(\left \vert \bm{k}_1+\bm{k}_2\right \vert \right)
+\tilde{C}_R\left(\left \vert \bm{k}_1+\bm{k}_3\right \vert \right)\right]
\right\rbrace \delta\left({\bm k}_1+{\bm k}_2+{\bm k}_3+{\bm k}_4\right)
,
\\
\label{eq:source8}
\mathfrak{S}_8&=\frac{4\gamma}{\left(2\pi\right)^{3/2}}
\left\lbrace
P_{pv}\left(\bm{k}_1\right) P_{vv}\left(\bm{k}_2\right)
\left[\tilde{C}_R\left(\left \vert \bm{k}_1+\bm{k}_3\right \vert \right)+
\tilde{C}_R\left(\left \vert \bm{k}_1+\bm{k}_4\right \vert \right)\right]
\nonumber \right. \\ &  \left.
+ P_{vv}\left(\bm{k}_2\right)
\left[\tilde{C}_R\left(\left \vert \bm{k}_1+\bm{k}_2\right \vert 
\right)P_{pv}\left(\bm{k}_3\right)+
\tilde{C}_R\left(\left \vert \bm{k}_1+\bm{k}_3\right \vert \right)
P_{vp}\left(\bm{k}_3\right)\right]
\nonumber \right. \\ &  \left.
+P_{vp}\left(\bm{k}_4\right) P_{vv}\left(\bm{k}_2\right)
\left[\tilde{C}_R\left(\left \vert \bm{k}_1+\bm{k}_2\right \vert \right)+
\tilde{C}_R\left(\left \vert \bm{k}_1+\bm{k}_4\right \vert \right)\right]
\right\rbrace \delta\left({\bm k}_1+{\bm k}_2+{\bm k}_3+{\bm k}_4\right)
,
\\
\label{eq:source9}
\mathfrak{S}_9&=
\frac{4\gamma}{\left(2\pi\right)^{3/2}}
\left\lbrace
P_{pv}\left(\bm{k}_1\right) P_{vv}\left(\bm{k}_3\right)
\left[\tilde{C}_R\left(\left \vert \bm{k}_1+\bm{k}_2\right \vert\right)+
\tilde{C}_R\left(\left \vert \bm{k}_1+\bm{k}_4\right \vert \right)\right]
\nonumber \right. \\ & \left.
+P_{vv}\left(\bm{k}_3\right)
\left[\tilde{C}_R\left(\left \vert \bm{k}_1+\bm{k}_2\right \vert 
\right)P_{vp}\left(\bm{k}_2\right) +
\tilde{C}_R\left(\left \vert \bm{k}_1+\bm{k}_3\right \vert \right)
P_{pv}\left(\bm{k}_2\right) \right]
\nonumber \right. \\ & \left.
+
P_{vp}\left(\bm{k}_4\right) P_{vv}\left(\bm{k}_3\right)
\left[\tilde{C}_R\left(\left \vert \bm{k}_1+\bm{k}_3\right \vert \right)+
\tilde{C}_R\left(\left \vert \bm{k}_1+\bm{k}_4\right \vert \right)\right]
\right\rbrace \delta\left({\bm k}_1+{\bm k}_2+{\bm k}_3+{\bm k}_4\right)
,
\\
\label{eq:source10}
\mathfrak{S}_{10}&=\frac{4\gamma}{\left(2\pi\right)^{3/2}}
\left\lbrace
P_{pv}\left(\bm{k}_1\right) P_{vv}\left(\bm{k}_4\right)
\left[\tilde{C}_R\left(\left \vert \bm{k}_1+\bm{k}_2\right \vert\right)+
\tilde{C}_R\left(\left \vert \bm{k}_1+\bm{k}_3\right \vert \right)\right]
\nonumber \right. \\ & \left.
+P_{vv}\left(\bm{k}_4\right)
\left[\tilde{C}_R\left(\left \vert \bm{k}_1+\bm{k}_2\right \vert 
\right)P_{vp}\left(\bm{k}_2\right)+
\tilde{C}_R\left(\left \vert \bm{k}_1+\bm{k}_4\right \vert 
\right)P_{pv}\left(\bm{k}_2\right)\right]
\nonumber \right. \\ & \left.
+P_{vp}\left(\bm{k}_3\right) P_{vv}\left(\bm{k}_4\right)
\left[\tilde{C}_R\left(\left \vert \bm{k}_1+\bm{k}_3\right \vert \right)+
\tilde{C}_R\left(\left \vert \bm{k}_1+\bm{k}_4\right \vert \right)\right]
\right\rbrace \delta\left({\bm k}_1+{\bm k}_2+{\bm k}_3+{\bm k}_4\right)
,
\\
\label{eq:source11}
\mathfrak{S}_{11}&=\frac{4\gamma}{\left(2\pi\right)^{3/2}}
\left\lbrace
P_{pv}\left(\bm{k}_1\right) P_{pv}\left(\bm{k}_2\right)
\left[\tilde{C}_R\left(\left \vert \bm{k}_1+\bm{k}_3\right \vert \right)+
\tilde{C}_R\left(\left \vert \bm{k}_1+\bm{k}_4\right \vert \right)\right]
\nonumber \right. \\ & \left.
+P_{pv}\left(\bm{k}_1\right) 
\left[\tilde{C}_R\left(\left \vert \bm{k}_1+\bm{k}_2\right \vert 
\right)P_{pv}\left(\bm{k}_3\right)+
\tilde{C}_R\left(\left \vert \bm{k}_1+\bm{k}_4\right \vert 
\right)P_{vp}\left(\bm{k}_3\right)\right]
\nonumber \right. \\ & \left.
+
\left[\tilde{C}_R\left(\left \vert \bm{k}_1+\bm{k}_2\right \vert \right)
P_{pv}\left(\bm{k}_3\right)P_{vp}\left(\bm{k}_2\right) +
\tilde{C}_R\left(\left \vert \bm{k}_1+\bm{k}_3\right \vert 
\right)P_{vp}\left(\bm{k}_3\right)P_{pv}\left(\bm{k}_2\right) \right]
\nonumber \right. \\ & \left.
+P_{pv}\left(\bm{k}_1\right)P_{vp}\left(\bm{k}_4\right) 
\left[\tilde{C}_R\left(\left \vert \bm{k}_1+\bm{k}_2\right \vert \right)+
\tilde{C}_R\left(\left \vert \bm{k}_1+\bm{k}_3\right \vert \right) \right]
\nonumber \right. \\ & \left.
+P_{vp}\left(\bm{k}_4\right)
\left[\tilde{C}_R\left(\left \vert \bm{k}_1+\bm{k}_2\right \vert 
\right)P_{vp}\left(\bm{k}_2\right) +
\tilde{C}_R\left(\left \vert \bm{k}_1+\bm{k}_4\right \vert \right)
P_{pv}\left(\bm{k}_2\right) \right]
\nonumber \right. \\ & \left.
+
\left[\tilde{C}_R\left(\left \vert \bm{k}_1+\bm{k}_3\right \vert \right)
P_{vp}\left(\bm{k}_3\right)P_{vp}\left(\bm{k}_4\right) +
\tilde{C}_R\left(\left \vert \bm{k}_1+\bm{k}_4\right \vert \right)
P_{vp}\left(\bm{k}_4\right)P_{vp}\left(\bm{k}_3\right) \right]
\right\rbrace 
\nonumber \\ & \times
\delta\left({\bm k}_1+{\bm k}_2+{\bm k}_3+{\bm k}_4\right)
.
\end{align}

\subsection{Master equation for the trispectrum}
\label{subsec:mastertri}

Combining the first-order differential equation for the sixteen four-point correlators, one can obtain a sixteenth-order differential equation for $\left \langle 
\hat{v}_{{\bm k}_1}
\hat{v}_{{\bm k}_2}
\hat{v}_{{\bm k}_3}
\hat{v}_{{\bm k}_4}
\right\rangle_\uc$. In the equilateral configuration, where the
four vectors ${\bm k}_1$, ${\bm k}_2$, ${\bm k}_3$ and ${\bm k}_4$
have the same modulus, the order of that equation can be reduced. In such a case indeed, in the equations of the
previous section, we do not need to distinguish between $\omega^2(k_1)$, $\omega^2(k_2)$, $\omega^2(k_3)$ and $\omega^2(k_4)$. In the following, we will simply
denote those quantities by $\omega ^2$. Moreover, in the source terms,
the argument of the environmental correlation functions will differ
only because, in a losange, the modulus
$\left \vert {\bm k}_i+{\bm k}_j\right \vert$ depends on $\alpha $,
the semi angle at the top (the angle between $\bm{k}_1$ and
$\bm{k}_1+\bm{k}_2$). The restriction to equilateral configurations is therefore
a great technical simplification.

Since we are interested in the four-point correlation function of
curvature perturbations, we start with \Eq{eq:vvvv}. Then, the strategy is to differentiate this
expression and use the other equations of
\Sec{subsec:fourquadratic} in order to obtain a closed differential equation
in
$\left\langle \hat{v}_{{\bm k}_1} \hat{v}_{{\bm k}_2}\hat{v}_{{\bm
      k}_3}\hat{v}_{{\bm k}_4}\right\rangle_\uc$.
The very same method led to a third-order differential equation for
the power spectrum. Here, we will see that the process stops when the
fifth time derivative of
$\left\langle \hat{v}_{{\bm k}_1} \hat{v}_{{\bm k}_2}\hat{v}_{{\bm
      k}_3}\hat{v}_{{\bm k}_4}\right\rangle_\uc$
is considered. Differentiating \Eq{eq:vvvv}, one obtains
\begin{align}
\label{eq:d2vvvv}
\frac{\dd^2  }{\dd  \eta^2} \left\langle \hat{v}_{{\bm k}_1}
\hat{v}_{{\bm k}_2}\hat{v}_{{\bm k}_3}\hat{v}_{{\bm k}_4}\right\rangle_\uc &=
2\Bigl(\left\langle \hat{v}_{{\bm k}_1} \hat{v}_{{\bm k}_2}\hat{p}_{{\bm k}_3}
\hat{p}_{{\bm k}_4}\right\rangle_\uc
+\left\langle \hat{p}_{{\bm k}_1} \hat{v}_{{\bm k}_2}\hat{v}_{{\bm k}_3}
\hat{p}_{{\bm k}_4}\right\rangle_\uc
%\nonumber \\ &
+\left\langle \hat{p}_{{\bm k}_1} \hat{v}_{{\bm k}_2}\hat{p}_{{\bm k}_3}
\hat{v}_{{\bm k}_4}\right\rangle_\uc
\nonumber \\ & 
+\left\langle \hat{v}_{{\bm k}_1} \hat{p}_{{\bm k}_2}\hat{p}_{{\bm k}_3}
\hat{v}_{{\bm k}_4}\right\rangle_\uc
+\left\langle \hat{v}_{{\bm k}_1} \hat{p}_{{\bm k}_2}\hat{v}_{{\bm k}_3}
\hat{p}_{{\bm k}_4}\right\rangle_\uc
+\left\langle \hat{p}_{{\bm k}_1} \hat{p}_{{\bm k}_2}\hat{v}_{{\bm k}_3}
\hat{v}_{{\bm k}_4}\right\rangle_\uc\Bigr)
\nonumber \\ &
-4 \omega^2
\left\langle \hat{v}_{{\bm k}_1} \hat{v}_{{\bm k}_2}\hat{v}_{{\bm k}_3}
\hat{v}_{{\bm k}_4}\right\rangle_\uc,
\end{align}
where one has used \Eqs{eq:vvvp1}, (\ref{eq:vvvp2}),
(\ref{eq:vvvp3}) and~(\ref{eq:vvvp4}) to express the derivatives of
correlators containing three $\hat{v}_{{\bm k}_i}$ and one
$\hat{p}_{{\bm k}_j}$. Looking at those equations, we see that they
are sourceless which explains why the above equation does not contain
any source term. Then, we differentiate once more and this leads to 
\begin{align}
\label{eq:d3vvvv}
\frac{\dd^3  }{\dd  \eta^3} \left\langle \hat{v}_{{\bm k}_1}
\hat{v}_{{\bm k}_2}\hat{v}_{{\bm k}_3}\hat{v}_{{\bm k}_4}
\right\rangle_\uc &=
6\Bigl( \left\langle \hat{v}_{{\bm k}_1} \hat{p}_{{\bm k}_2}
\hat{p}_{{\bm k}_3}\hat{p}_{{\bm k}_4}\right\rangle_\uc
+\left\langle \hat{p}_{{\bm k}_1} \hat{v}_{{\bm k}_2}
\hat{p}_{{\bm k}_3}\hat{p}_{{\bm k}_4}\right\rangle_\uc
+\left\langle \hat{p}_{{\bm k}_1} \hat{p}_{{\bm k}_2}
\hat{v}_{{\bm k}_3}\hat{p}_{{\bm k}_4}\right\rangle_\uc
\nonumber \\ &
+\left\langle \hat{p}_{{\bm k}_1} \hat{p}_{{\bm k}_2}
\hat{p}_{{\bm k}_3}\hat{v}_{{\bm k}_4}\right\rangle_\uc\Bigr)
-6\omega^2\Bigl(
\left\langle \hat{v}_{{\bm k}_1} \hat{v}_{{\bm k}_2}
\hat{p}_{{\bm k}_3}\hat{v}_{{\bm k}_4}\right\rangle_\uc
+\left\langle \hat{v}_{{\bm k}_1} \hat{v}_{{\bm k}_2}
\hat{v}_{{\bm k}_3}\hat{p}_{{\bm k}_4}\right\rangle_\uc
\nonumber \\ &
+\left\langle \hat{p}_{{\bm k}_1} \hat{v}_{{\bm k}_2}
\hat{v}_{{\bm k}_3}\hat{v}_{{\bm k}_4}\right\rangle_\uc
+ \left\langle \hat{v}_{{\bm k}_1} \hat{p}_{{\bm k}_2}
\hat{v}_{{\bm k}_3}\hat{v}_{{\bm k}_4}\right\rangle_\uc\Bigr)
-4\frac{\dd}{\dd\eta}\left(\omega^2\left\langle \hat{v}_{{\bm k}_1}
\hat{v}_{{\bm k}_2}\hat{v}_{{\bm k}_3}\hat{v}_{{\bm k}_4}
\right\rangle_\uc \right)
\nonumber \\&
+ 2\sum_{i=1}^6 \mathfrak{S}_i, 
\end{align}
where, this time, one has used \Eqs{eq:vvpp1}, (\ref{eq:vvpp2}),
(\ref{eq:vvpp3}), (\ref{eq:vvpp4}), (\ref{eq:vvpp5})
and~(\ref{eq:vvpp6}). One notices the appearance, for the first time,
of source terms. Then, one needs to differentiate once more, leading to 
the following expression
\begin{align}
\label{eq:d4vvvv}
\frac{\dd^4  }{\dd  \eta^4} \left\langle \hat{v}_{{\bm k}_1}
\hat{v}_{{\bm k}_2}\hat{v}_{{\bm k}_3}\hat{v}_{{\bm k}_4}\right\rangle_\uc 
&=
24 \left\langle \hat{p}_{{\bm k}_1}
\hat{p}_{{\bm k}_2}\hat{p}_{{\bm k}_3}\hat{p}_{{\bm k}_4}\right\rangle_\uc
-24\omega^2\Bigl(
\left\langle \hat{v}_{{\bm k}_1} \hat{p}_{{\bm k}_2}\hat{p}_{{\bm k}_3}
\hat{v}_{{\bm k}_4}\right\rangle_\uc
+
\left\langle \hat{v}_{{\bm k}_1} \hat{p}_{{\bm k}_2}\hat{v}_{{\bm k}_3}
\hat{p}_{{\bm k}_4}\right\rangle_\uc
\nonumber \\ &
+
\left\langle \hat{v}_{{\bm k}_1} \hat{v}_{{\bm k}_2}\hat{p}_{{\bm k}_3}
\hat{p}_{{\bm k}_4}\right\rangle_\uc
+
\left\langle \hat{p}_{{\bm k}_1} \hat{v}_{{\bm k}_2}\hat{p}_{{\bm k}_3}
\hat{v}_{{\bm k}_4}\right\rangle_\uc
+
\left\langle \hat{p}_{{\bm k}_1} \hat{v}_{{\bm k}_2}\hat{v}_{{\bm k}_3}
\hat{p}_{{\bm k}_4}\right\rangle_\uc
\nonumber \\ &
+
\left\langle \hat{p}_{{\bm k}_1} \hat{p}_{{\bm k}_2}
\hat{v}_{{\bm k}_3}\hat{v}_{{\bm k}_4}\right\rangle_\uc\Bigr)
-6{\omega^2}^\prime 
\Bigl(
\left\langle \hat{v}_{{\bm k}_1} \hat{v}_{{\bm k}_2}\hat{p}_{{\bm k}_3}
\hat{v}_{{\bm k}_4}\right\rangle_\uc
+
\left\langle \hat{v}_{{\bm k}_1} \hat{v}_{{\bm k}_2}\hat{v}_{{\bm k}_3}
\hat{p}_{{\bm k}_4}\right\rangle_\uc
\nonumber \\ &
+
\left\langle \hat{p}_{{\bm k}_1} \hat{v}_{{\bm k}_2}
\hat{v}_{{\bm k}_3}\hat{v}_{{\bm k}_4}\right\rangle_\uc
+
\left\langle \hat{v}_{{\bm k}_1} \hat{p}_{{\bm k}_2}
\hat{v}_{{\bm k}_3}\hat{v}_{{\bm k}_4}\right\rangle_\uc\Bigr)
+24 \omega^4\left\langle \hat{v}_{{\bm k}_1}
\hat{v}_{{\bm k}_2}\hat{v}_{{\bm k}_3}\hat{v}_{{\bm k}_4}\right\rangle_\uc
\nonumber \\ &
-4\frac{\dd^2}{\dd\eta^2}\left( \omega^2\left\langle \hat{v}_{{\bm k}_1}
\hat{v}_{{\bm k}_2}\hat{v}_{{\bm k}_3}\hat{v}_{{\bm k}_4}\right\rangle_\uc \right)
+2\sum_{i=1}^6 \mathfrak{S}_i' +6 \sum_{i=7}^{10}\mathfrak{S}_i,
\end{align}
where \Eqs{eq:vppp1}, (\ref{eq:vppp2}, (\ref{eq:vppp3})
and~(\ref{eq:vppp4}) but also~(\ref{eq:vvvp1}), (\ref{eq:vvvp2}),
(\ref{eq:vvvp3}), (\ref{eq:vvvp4}) have been used. Finally, one steps
remains to be completed and one has to differentiate a last time. One
obtains
\begin{align}
\label{eq:d5vvvv}
\frac{\dd^5  }{\dd  \eta^5} \left\langle \hat{v}_{{\bm k}_1}
\hat{v}_{{\bm k}_2}\hat{v}_{{\bm k}_3}\hat{v}_{{\bm k}_4}\right\rangle_\uc &=
-96 \omega^2\Bigl(
\left\langle \hat{p}_{{\bm k}_1}\hat{p}_{{\bm k}_2}\hat{p}_{{\bm k}_3}
\hat{v}_{{\bm k}_4}\right\rangle_\uc
+\left\langle \hat{p}_{{\bm k}_1}\hat{p}_{{\bm k}_2}\hat{v}_{{\bm k}_3}
\hat{p}_{{\bm k}_4}\right\rangle_\uc
+\left\langle \hat{p}_{{\bm k}_1}\hat{v}_{{\bm k}_2}\hat{p}_{{\bm k}_3}
\hat{p}_{{\bm k}_4}\right\rangle_\uc
\nonumber \\ &
+\left\langle \hat{v}_{{\bm k}_1}\hat{p}_{{\bm k}_2}\hat{p}_{{\bm k}_3}
\hat{p}_{{\bm k}_4}\right\rangle_\uc
\Bigr)
+72\omega^4\Bigl(
\left\langle \hat{v}_{{\bm k}_1}\hat{p}_{{\bm k}_2}\hat{v}_{{\bm k}_3}
\hat{v}_{{\bm k}_4}\right\rangle_\uc
+\left\langle \hat{v}_{{\bm k}_1}\hat{v}_{{\bm k}_2}\hat{p}_{{\bm k}_3}
\hat{v}_{{\bm k}_4}\right\rangle_\uc
\nonumber \\ &
+\left\langle \hat{v}_{{\bm k}_1}\hat{v}_{{\bm k}_2}\hat{v}_{{\bm k}_3}
\hat{p}_{{\bm k}_4}\right\rangle_\uc
+\left\langle \hat{p}_{{\bm k}_1}\hat{v}_{{\bm k}_2}\hat{v}_{{\bm k}_3}
\hat{v}_{{\bm k}_4}\right\rangle_\uc
\Bigr)
-72\omega\omega' \Bigl(
\left\langle \hat{v}_{{\bm k}_1}\hat{p}_{{\bm k}_2}
\hat{p}_{{\bm k}_3}\hat{v}_{{\bm k}_4}\right\rangle_\uc
\nonumber \\ &
+ \left\langle \hat{v}_{{\bm k}_1}\hat{p}_{{\bm k}_2}
\hat{v}_{{\bm k}_3}\hat{p}_{{\bm k}_4}\right\rangle_\uc
+ \left\langle \hat{v}_{{\bm k}_1}\hat{v}_{{\bm k}_2}
\hat{p}_{{\bm k}_3}\hat{p}_{{\bm k}_4}\right\rangle_\uc
+\left\langle \hat{p}_{{\bm k}_1}\hat{v}_{{\bm k}_2}
\hat{p}_{{\bm k}_3}\hat{v}_{{\bm k}_4}\right\rangle_\uc
\nonumber \\ &
+ \left\langle \hat{p}_{{\bm k}_1}\hat{v}_{{\bm k}_2}
\hat{v}_{{\bm k}_3}\hat{p}_{{\bm k}_4}\right\rangle_\uc
+\left\langle \hat{p}_{{\bm k}_1}\hat{p}_{{\bm k}_2}
\hat{v}_{{\bm k}_3}\hat{v}_{{\bm k}_4}\right\rangle_\uc
\Bigr)
-6\left(\omega^2\right)''\Bigl(
 \left\langle \hat{p}_{{\bm k}_1}\hat{v}_{{\bm k}_2}
\hat{v}_{{\bm k}_3}\hat{v}_{{\bm k}_4}\right\rangle_\uc
\nonumber \\ &
+\left\langle \hat{v}_{{\bm k}_1}\hat{p}_{{\bm k}_2}
\hat{v}_{{\bm k}_3}\hat{v}_{{\bm k}_4}\right\rangle_\uc
+ \left\langle \hat{v}_{{\bm k}_1}\hat{v}_{{\bm k}_2}
\hat{p}_{{\bm k}_3}\hat{v}_{{\bm k}_4}\right\rangle_\uc
+\left\langle \hat{v}_{{\bm k}_1}\hat{v}_{{\bm k}_2}
\hat{v}_{{\bm k}_3}\hat{p}_{{\bm k}_4}\right\rangle_\uc
\Bigr)
\nonumber \\ &
%%%%%
+144\omega^3 \omega'
\left\langle \hat{v}_{{\bm k}_1}\hat{v}_{{\bm k}_2}\hat{v}_{{\bm k}_3}
\hat{v}_{{\bm k}_4}\right\rangle_\uc
+24\omega^4\frac{\dd}{\dd\eta}\left\langle \hat{v}_{{\bm k}_1}
\hat{v}_{{\bm k}_2}\hat{v}_{{\bm k}_3}\hat{v}_{{\bm k}_4}\right\rangle_\uc
\nonumber \\ &
-4\frac{\dd^3}{\dd\eta^3}\left(\omega^2 \left\langle \hat{v}_{{\bm k}_1}
\hat{v}_{{\bm k}_2}\hat{v}_{{\bm k}_3}\hat{v}_{{\bm k}_4}\right\rangle_\uc\right)
+24 \mathfrak{S}_{11}
+2\sum_{i=1}^6 \mathfrak{S}_i'' + 6 \sum_{i=7}^{10}\mathfrak{S}_i'
\nonumber \\ &
-24 \omega^2\sum_{i=1}^6 \mathfrak{S}_i,
\end{align}
where all the equations for the derivative of the four-point
correlators have been used. The process stops at this stage because
combining \Eqs{eq:vvvv}, (\ref{eq:d2vvvv}), (\ref{eq:d3vvvv}),
(\ref{eq:d4vvvv}) and~(\ref{eq:d5vvvv}), the terms arrange themselves
such that only the correlator
$\left\langle \hat{v}_{{\bm k}_1}\hat{v}_{{\bm k}_2}\hat{v}_{{\bm
      k}_3}\hat{v}_{{\bm k}_4}\right\rangle_\uc$ appears. This leads to
\begin{align}
\label{eq:masterappendix}
\Bigl\lbrace
\frac{\dd^5}{\dd\eta^5}
& +20\omega^2\frac{\dd^3}{\dd\eta^3}
+60\omega\omega'\frac{\dd^2}{\dd\eta^2}
+\left[64\omega^4+18\left(\omega^2\right)''\right]\frac{\dd}{\dd\eta}
\nonumber \\ &
+\left[128 \omega^3\omega'+4\left(\omega^2\right)'''\right]
\Bigr\rbrace
\left\langle \hat{v}_{{\bm k}_1}\hat{v}_{{\bm k}_2}\hat{v}_{{\bm k}_3}\hat{v}_{{\bm k}_4}\right\rangle_\uc
= 
2\sum_{i=1}^6\left(\mathfrak{S}_i''+4\omega^2\mathfrak{S}_i\right)
+6\sum_{i=7}^{10}\mathfrak{S}_i'+24\mathfrak{S}_{11},
\end{align}
which corresponds to \Eq{eq:mastereq} used in the main text. Using the form
given above for the source functions, see \Eqs{eq:source1},
(\ref{eq:source2}), (\ref{eq:source3}), (\ref{eq:source4}),
(\ref{eq:source5}), (\ref{eq:source6}), (\ref{eq:source7}),
(\ref{eq:source8}), (\ref{eq:source9}), (\ref{eq:source10})
and~(\ref{eq:source11}), one has
\begin{align}
\label{eq:sumsource1}
\sum_{i=1}^6\mathfrak{S}_i &= \frac{16\gamma}{\left(2\pi\right)^{{3/2}}}
P_{vv}^2(k) 
\left[\tilde{C}_R\left(2k \cos\alpha\right)
+\tilde{C}_R\left(2k \sin\alpha\right)+\tilde{C}_R(0)\right]
\delta\left(\bm{k}_1+\bm{k}_2+\bm{k}_3+\bm{k}_4\right),
\\
\label{eq:sumsource2}
\sum_{i=7}^{10}\mathfrak{S}_i &= \frac{16\gamma}{\left(2\pi\right)^{{3/2}}}
P_{vv}(k)
\left[P_{vp}(k)+P_{pv}(k)\right] \left[\tilde{C}_R\left(2k \cos\alpha\right)
+\tilde{C}_R\left(2k \sin\alpha\right)+\tilde{C}_R(0)\right]
\nonumber \\ & \times 
\delta\left(\bm{k}_1+\bm{k}_2+\bm{k}_3+\bm{k}_4\right),
\\
\label{eq:sumsource3}
\mathfrak{S}_{11} &= \frac{4\gamma}{\left(2\pi\right)^{{3/2}}}
\left[P_{vp}(k)+P_{pv}(k)\right]^2 \left[\tilde{C}_R\left(2k \cos\alpha\right)
+\tilde{C}_R\left(2k \sin\alpha\right)+\tilde{C}_R(0)\right]
\nonumber \\ & \times
\delta\left(\bm{k}_1+\bm{k}_2+\bm{k}_3+\bm{k}_4\right).
\end{align}
Notice that if the losange is a square, then $\alpha =\pi/4$ and one
checks that the modulus of vectors ${\bm k}_i+{\bm k}_j$ appearing in
the argument of the correlation function are all equal. Finally,
looking at \Eq{eq:masterappendix}, we see that one must define
the ``total'' source function as
\begin{align}
\mathfrak{S}(\eta, {\bm k},\alpha)\equiv 2\sum_{i=1}^6\left(\mathfrak{S}_i''
+4\omega^2\mathfrak{S}_i\right)
+6\sum_{i=7}^{10}\mathfrak{S}_i'+24\mathfrak{S}_{11},
\label{eq:trispectrum:equilateral:quadratic:eom}
\end{align}
the expression of which, using \Eqs{eq:sumsource1}-(\ref{eq:sumsource3}) together with \Eqs{eq:vvquadratic}-(\ref{eq:ppquadratic}), exactly lead to
\Eq{eq:sourcetri} used in the main text.

\subsection{Solution to the master equation}
\label{subsec:solmastertri}

Our goal in this section is to solve \Eq{eq:mastereq} [or
\Eq{eq:masterappendix}]. Since we know that the solution is
given by \Eq{eq:soltri}, the problem is in fact to calculate
the integral present in the solution. For our purpose, it is
enough to perform this calculation in de Sitter since slow-roll corrections would only affect the constraints we derive in a negligible way. In that case,
$\omega^2=k^2-2/\eta^2$ (we recall that $\eta$ is the conformal time)
and the mode function is simply given by 
\begin{align} 
\label{eq:vk:deSitter}
v_{\bm k}(\eta) =\frac{\ee^{-i k \eta}}{\sqrt{2k}}
\left(1-\frac{i}{k\eta}\right)\, .  
\end{align} 
This allows us to calculate the two functions appearing in the integrand 
of \Eq{eq:soltri}, namely $\mathfrak{S}(\eta, {\bm k},\alpha)$ 
and $\Ima\left[v_{\bm k}(\eta')v_{\bm k}^*(\eta)\right]$.

Let us start with the later since it is obviously very
simple. Using \Eq{eq:vk:deSitter}, one has
\begin{align}
\label{eq:Imappendix}
\Ima\left[v_{\bm k}(\eta')v_{\bm k}^*(\eta)\right]=\frac{
\cos\left[k\left(\eta-\eta'\right)\right]}{2k}
\left(\frac{1}{k\eta}-\frac{1}{k\eta'}\right)
+\frac{\sin\left[k\left(\eta-\eta'\right)\right]}{2k}
\left(1+\frac{1}{k^2\eta\eta'}\right)\, .
\end{align}

The calculation of the source~(\ref{eq:sourcetri}) is clearly more
involved. We recall that the $\gamma$ parameter depends on time according to
$\gamma =\gamma _*(a/a_*)^p$ and that the correlation function of the environment is given by
\Eq{eq:Ck:appr}. It follows that the amplitude of the source
is determined by the dimensionless parameter
$\sigma_\gamma=\bar{C}_R\lE^3\gamma_*/a_*^3$. Since we
use the de Sitter solution, one can also evaluate $P_{vv}(k)$,
$P_{vp}(k)+P_{pv}(k)$ and $P_{pp}(k)$ exactly and they read
\begin{align}
P_{vv}(k)&=\frac{1}{2k}\left[1+\frac{1}{\left(-k\eta\right)^2}\right], \quad 
P_{vp}(k)+P_{pv}(k)=\frac{1}{\left(-k\eta\right)^3},\\
P_{pp}(k)&=\frac{k}{2}\left[1-\frac{1}{\left(-k\eta\right)^2}
+\frac{1}{\left(-k\eta\right)^4}\right].
\end{align} 
Inserting these results in \Eq{eq:sourcetri}, one obtains the following 
expression for the source term
\begin{align}
\label{eq:sourceappendix}
\mathfrak{S}(\eta, {\bm k},\alpha)
&=\frac{2\sigma_{\gamma}}{3\pi^2}\left(\frac{k}{k_*}\right)^{p-3}
\biggr\lbrace {\cal F}_1(-k\eta)\left[
\Theta\left(-2k\eta H\lE\cos \alpha\right)
+\Theta\left(-2k\eta H\lE\sin\alpha\right)+1\right]
\nonumber \\ &
+\frac{{\cal F}_2(-k\eta)}{k}\left[
\delta\left(\eta+\frac{1}{2k\cos \alpha H\lE}\right)
+\delta\left(\eta+\frac{1}{2k\sin \alpha H\lE}\right)\right]
\nonumber \\ &
+\frac{{\cal F}_3(-k\eta)}{k^2}\left[
\delta'\left(\eta+\frac{1}{2k\cos \alpha H\lE }\right)
+\delta'\left(\eta+\frac{1}{2k\sin \alpha H\lE }\right)\right]
\biggr\rbrace 
\nonumber \\ & \times
\delta\left(\bm{k}_1+\bm{k}_2+\bm{k}_3+\bm{k}_4\right),
\end{align}
where one has used the de Sitter scale factor, $a=-1/(H\eta)$, and where we have defined
\begin{align}
{\cal F}_1(u) &= 
u^{-3-p}
\left[8u^6+2(p-2)(p-3)uu^4-24u^2+4p(p+2)u^2+2p^2
+18p+36\right], \\
{\cal F}_2(u) &= 8u^{-2-p}
\left[(p-3)u^4+(2p+1)u^2+p+4\right],\\
{\cal F}_3(u) &= 
8u^{3-p}\left(1+u^{-2}\right).
\end{align}
The appearance of the Dirac delta function and of its derivative is of
course related to the fact that the source term contains the
derivative and the second derivative of the environmental correlation
function with respect to time.

At this stage, in principle, all we have to do is to insert
\Eqs{eq:Imappendix} and~(\ref{eq:sourceappendix}) into
\Eq{eq:soltri}, which gives three contributions, respectively
proportional to ${\cal F}_1$, ${\cal F}_2$ and ${\cal
  F}_3$. Explicitly, the first contribution reads
\begin{align}
\left\langle  \hat{v}_{{\bm k}_1}\hat{v}_{{\bm k}_2}\hat{v}_{{\bm k}_3}
\hat{v}_{{\bm k}_4}\right\rangle_\uc^{(1)}  &=
\frac{\sigma_{\gamma}}{36\pi^2k^5}
\left(\frac{k}{k_*}\right)^{p-3}
\int ^{k\eta}_{-\infty}
\left[\cos\left(k\eta-z\right)
\left(\frac{1}{k\eta}-\frac{1}{z}\right)
\right. \nonumber \\ & \left.
+\sin\left(k\eta-z\right)
\left(1+\frac{1}{k\eta z}\right)
\right]^4{\cal F}_1(-z)
\left[
\Theta\left(-2z H\lE \cos \alpha\right)
\right. \nonumber \\ & \left.
+\Theta\left(-2z H\lE \sin\alpha\right)+1\right]
{\rm d}z \, \delta\left(\bm{k}_1+\bm{k}_2+\bm{k}_3+\bm{k}_4\right).
\end{align}
A few remarks are in order at this point. First, the infinite lower bound of the integral
becomes finite for the two terms proportional to the
$\Theta$ functions [and respectively given by
$-(2H\lE \cos \alpha)^{-1}$ and
$-(2H\lE \sin \alpha)^{-1}$], which insures that the
corresponding integrals are finite. For the third term, proportional
to one, convergence is not clear a priori but one can regulate the integral
setting the upper bound to $k\eta_{_\mathrm{IR}}$,
$\eta_{_\mathrm{IR}}$ being \eg the time at the beginning of
inflation. Let us however notice that this term comes from evaluating the environment correlation function for a vanishing momentum. Usually, such terms are viewed as being re-absorbable in the background and thus, on this ground, are often ignored. This is what is done in the results presented in the main text. However, if they were to be included, this would drastically improve the constraints in the case $p<4$. Therefore, discarding these infrared effects in this context yields to conservative constraints. Admittedly, the status of such infrared effects, which is already a matter of debates for the power spectrum, is even more acute for higher correlation functions since the dependence on the infrared cutoff is a power law in that case while it is only logarithmic for the two-point correlation functions~\cite{Martin:2018zbe}. 

Second, the above integral remains complicated but we are
interested in the large-scale limit only, namely
$k\eta \rightarrow 0$. Third, in this limit, the behaviour of the
correlators can be obtained by identifying the region in the integration domain from where the integral receives its main contribution. When $p<4$, the integral is dominated by the neighbourhood
of its lower bound $z_{\mathrm{min}}$. Expanding the integrand in the
limit $-k\eta\ll 1$ and $-z\gg 1$, one obtains that each of the three
terms (the two terms proportional to $\Theta$ and the one proportional to
$1$) are equal to
\begin{align}
\left\langle  \hat{v}_{{\bm k}_1}\hat{v}_{{\bm k}_2}\hat{v}_{{\bm k}_3}
\hat{v}_{{\bm k}_4}\right\rangle_\uc^{(1)}  \supset \frac{2\sigma_\gamma}
{9k^5\pi^2}\left(\frac{k}{k_*}\right)^{p-3}
\frac{3}{8(4-p)}\frac{\left(-z_{\mathrm{min}}\right)^{4-p}}
{(-k\eta)^4}\delta\left(\bm{k}_1+\bm{k}_2+\bm{k}_3+\bm{k}_4\right),
\end{align}
where $z_{\mathrm{min}}$ is either
$-(2H\lE \cos \alpha)^{-1}$,
$-(2H\lE \sin \alpha)^{-1}$ or
$k\eta_{_\mathrm{IR}}$. The case $p=4$ is singular
but can be worked out and one finds
\begin{align}
\left\langle  \hat{v}_{{\bm k}_1}\hat{v}_{{\bm k}_2}\hat{v}_{{\bm k}_3}
\hat{v}_{{\bm k}_4}\right\rangle_\uc^{(1)}  \supset \frac{2\sigma_\gamma}
{9k^5\pi^2}\frac{k}{k_*}
\frac{3}{8(-k\eta)^4}\left[\gamma _{_\mathrm{E}}+\ln \left(-4z_{\mathrm{min}}\right)\right]\delta\left(\bm{k}_1
+\bm{k}_2+\bm{k}_3+\bm{k}_4\right),
\end{align}
where $\gamma _{_\mathrm{E}}$ is the Euler constant.

If $4<p<6$, the main contribution to the integral comes from the
neighbourhood of $z\sim -1$ and no expansion is available. In that
case, the result can be worked out only up to an overall order of one prefactor,
\begin{align}
\left\langle  \hat{v}_{{\bm k}_1}\hat{v}_{{\bm k}_2}\hat{v}_{{\bm k}_3}
\hat{v}_{{\bm k}_4}\right\rangle_\uc^{(1)}\supset \frac{2\sigma_\gamma}
{9k^5\pi^2}\left(\frac{k}{k_*}\right)^{p-3}
\frac{\mathcal{O}(1)}{(k\eta)^4}\delta\left(\bm{k}_1
+\bm{k}_2+\bm{k}_3+\bm{k}_4\right)\, .
\end{align}
The $\mathcal{O}(1)$ constant, which depends on $p$, can be calculated analytically for half-integer values of $p$ as pointed out in footnote~\ref{footnote:interpolation}. For $p=5$ it is $7\pi/20\simeq 1.0996$, for $p=9/2$ it is $103\sqrt{\pi/2}/90 \simeq 1.4344$, and for
$p=11/2$ it is $194\sqrt{2\pi}/385\simeq 1.2631$. This confirms that
this constant is indeed of order one and thus its precise value does not matter
much.

The case $p=6$ is singular but one can show that the dominant contribution to the correlator reads
\begin{align}
\left\langle  \hat{v}_{{\bm k}_1}\hat{v}_{{\bm k}_2}\hat{v}_{{\bm k}_3}
\hat{v}_{{\bm k}_4}\right\rangle_\uc^{(1)}\supset \frac{2\sigma_\gamma}
{9k^5\pi^2}\left(\frac{k}{k_*}\right)^{3}\frac{1}{3\left(k\eta\right)^4}
\left[3-\gamma_{{}_\mathrm{E}}-\ln\left(-4k\eta\right)\right]
\delta\left(\bm{k}_1
+\bm{k}_2+\bm{k}_3+\bm{k}_4\right).
\end{align}
Finally remains the case $p>6$ where the integral is dominated by its
upper bound. Expanding the integrand in the limit
$-k\eta\ll 1$ and $-z\ll 1$, one finds 
\begin{align}
\left\langle  \hat{v}_{{\bm k}_1}\hat{v}_{{\bm k}_2}\hat{v}_{{\bm k}_3}
\hat{v}_{{\bm k}_4}\right\rangle_\uc^{(1)}\supset \frac{2\sigma_\gamma}
{9k^5\pi^2}\left(\frac{k}{k_*}\right)^{p-3}
\frac{6}{(p-6)(p-3)p}\left(-k\eta\right)^{2-p}\delta\left(\bm{k}_1
+\bm{k}_2+\bm{k}_3+\bm{k}_4\right)\, .
\end{align}
This completes our calculation of the first term in \Eq{eq:sourceappendix}, \ie the one proportional to $\mathcal{F}_1$.

The second and third terms to the four-point correlator can be
worked out in the same way. In fact, the calculation is easier since the integral is trivially performed via the Dirac delta
function and its derivative in those terms. It is easy to show that
they always lead to subdominant contributions that can, therefore, be
safely neglected. These considerations give rise to the formulas given in \Sec{subsec:paramgnl}.

\bibliographystyle{JHEP}
\bibliography{Lindblad_NG}
\end{document}